\documentclass[twocolumn,resetfootnotes]{aastex7}
\usepackage{amsmath}
\usepackage{empheq}

\usepackage{float} 
\revised{\today}

\usepackage{lineno}

\submitjournal{ApJ}
\def\avg#1{\langle#1\rangle}

\newcommand{\Em}{\dot{E}_{\rm max}}
\newcommand{\gv}[1]{\ensuremath{\mbox{\boldmath$ #1 $}}} \newcommand{\pdtext}[2]{\partial #1/\partial #2}
\usepackage{letltxmacro,amsmath}
\LetLtxMacro{\originaleqref}{\eqref}
\renewcommand{\eqref}{Eq.~\originaleqref}

\defcitealias{2025SoPh..300....5W}{WS25}
\defcitealias{2019ApJ...884...68K}{KL19}

\shortauthors{A. Stricklan, T. Waters, J. Klimchuk}
\shorttitle{Stability Analysis of Cooling Functions}

\begin{document}

\title{On the stability analysis of astrophysical cooling functions}
\author[0000-0001-7751-8885]{Amanda Stricklan}
\affiliation{Los Alamos National Laboratory, Los Alamos, New Mexico 87545, USA}
\affiliation{Department of Astronomy, New Mexico State University \ P.O. Box 30001, MSC 4500 \ Las Cruces, NM 88003-8001, USA }
\email[show]{astricklan@lanl.gov}  
\author[0000-0002-5205-9472]{Tim Waters}
\affiliation{Los Alamos National Laboratory, Los Alamos, New Mexico 87545, USA}
\affiliation{Center for Theoretical Astrophysics, Los Alamos National Laboratory, Los Alamos, NM, USA}
\email{waters@lanl.gov}  
\author[0000-0003-2255-0305]{James Klimchuk}
\affiliation{Heliophysics Science Division NASA Goddard Space Flight Center 8800 Greenbelt Road Greenbelt, MD 20771, USA}
\email{James.A.Klimchuk@nasa.gov} 

\begin{abstract} 
To model the temperature evolution of optically thin astrophysical environments at MHD scales, radiative and collisional cooling rates are typically either pre-tabulated or fit into a functional form and then input into MHD codes as a radiative loss function.
Thermal balance requires estimates of the analogous heating rates, which are harder to calculate, and due to uncertainties in the underlying dissipative heating processes, these rates are often simply parameterized.  
The resulting net cooling function defines an equilibrium curve that varies with density and temperature.  
Such cooling functions can make the gas prone to thermal instability (TI), which will cause departures from equilibrium.  
There has been no systematic study of thermally unstable parameter space for nonequilibrium states.
Motivated by our recent finding that there is a related linear instability, catastrophic cooling instability, that can dominate over TI, here we carry out such a study. 
We show that Balbus' instability criteria for TI can be used to define a critical cooling rate, $\Lambda_c$, that permits a nonequilibrium analysis of cooling functions through the mapping of TI zones.  
We furthermore extend Balbus' criteria to account for thermal conduction.  
Upon applying a $\Lambda_c$-based stability analysis to coronal loop simulations, we find that loops undergoing periodic episodes of  coronal rain formation are linearly unstable to catastrophic cooling instability, while TI is stabilized by thermal conduction.
\end{abstract}


\section{Introduction}
The radiative and collisional heating and cooling rates of both neutral and ionized gas have been extensively studied within the astrophysics and solar physics communities \citep[for a textbook account, see e.g.][]{2003adu..book.....D, 2004psci.book.....A, 2014masu.book.....P, 2021iim..book.....R}. 
Estimates of these rates are essential inputs for MHD codes capable of modeling condensation phenomena. 
The common practice is currently to calculate these rates up front as a function of temperature and density using plasma codes with atomic databases such as CHIANTI \citep{1997A&AS..125..149D} and APED/APEC \citep{2001ApJ...556L..91S}, MAPPINGS \citep{1993ApJS...88..253S}, SPEX \citep{1996uxsa.conf..411K}, XSTAR \citep{2001ApJS..133..221K}, and Cloudy \citep{2017RMxAA..53..385F}.
In studies of the interstellar medium (ISM), even the most sophisticated efforts to model heating and cooling processes for hydrogen self-consistently using a chemistry network \citep[e.g.][]{1999ApJ...524..923N, 2007ApJ...659.1317G, 2010MNRAS.404....2G, 2017ApJ...843...38G, 2023ApJS..268...42G} still rely on tabulated cooling functions for incorporating rates from heavier elements \citep[e.g.][]{2014MNRAS.440.3100W, 2023ApJS..264...10K, 2024MNRAS.528..255R}.
\par

The phase diagrams associated with ISM cooling functions typically reveal that there can be multi-temperature solutions at the same gas pressure \citep[see, for example,][]{1992pavi.book.....S}.  
The well known linear instability associated with these cooling functions, thermal instability (TI), can serve as the mechanism causing such `multiphase' solutions to appear when solving the MHD equations.  
The overall cooling and heating rates are sensitive to the elemental abundances and to the radiative and collisional processes taking place \citep[e.g.] []{2011ApJ...737...27X, 2015AdSpR..56.2738M, 2024A&A...688A.145J}, but for a given parcel of gas cooling at a specified rate, the stability analysis of the parcel is rather simple, depending on only two thermodynamic derivatives \citep{1965ApJ...142..531F}.  
\par

The nonlinear outcome of TI is to trigger catastrophic cooling, a runaway radiative cooling process in which gas cools rapidly to a new thermal equilibrium state.
In his comprehensive study on the linear theory of TI, \citet{1965ApJ...142..531F} showed that the eigenmodes found by \citet{1953ApJ...117..431P}, in the first paper on the subject, are not self-consistent solutions to the equations of gas dynamics because they were derived from the heat equation alone.  
However, he neglected to point out that both his and Parker's eigenmodes agree in the zero wavenumber limit where the evolution becomes truly isochoric.  
We recently showed that this special eigenmode\textemdash because it comes from the heat equation alone\textemdash is actually the thermal mode of an entirely separate instability \citep[see][hereafter referred to as \citetalias{2025SoPh..300....5W}]{2025SoPh..300....5W}.  
After presenting results of a very general nature about TI, we will turn our attention to the relevance of this catastrophic cooling instability to solving a longstanding problem regarding so-called thermal nonequilibrium (TNE), a cyclic process commonly observed in global coronal loop simulations \citep[e.g.][]{1991ApJ...378..372A, 1999ApJ...512..985A, 2010ApJ...714.1239K, 2017ApJ...835..272F, 2018ApJ...855...52F, 2019A&A...625A.149J, 2020PPCF...62a4016A, 2021ApJ...916..115S, 2022A&A...658A..71P, 2024ApJ...976..226S, 2025ApJ...978...94Y}.
\par

We mentioned both ISM and coronal loop modeling to point out a gap in how these research fields, which together encompass the majority of the early TI literature, have traditionally approached the stability analysis of TI.
The ISM community has long utilized phase diagrams to identify thermally unstable parameter space \citep[e.g.][]{2000ApJ...540..271V, 2002ApJ...569L.127K, 2003ApJ...587..278W, 2005AIPC..784..318I, 2013ApJ...779...48K, 2014A&A...567A..16S}. 
This type of analysis is rarely used in coronal loop modeling, not even in the study of prominences and coronal rain where TI is often invoked as the underlying formation process \citep[see the reviews by][]{2015ASSL..415..237K, 2020RAA....20..166C,2020PPCF...62a4016A,2022FrASS...920116A}.
This is likely because numerical models often utilize a position-dependent heating term that would make phase diagrams functions of height and hence difficult to analyze \citep[e.g.][]{2019ApJ...884...68K, 2020A&A...639A..20K, 2022A&A...668A..47B, 2022ApJ...926..216L}. 
Instead, such studies typically only show the cooling curve, which is a plot of the radiative loss function versus temperature \citep[e.g.][]{2020A&A...639A..20K, 2021SoPh..296..102W, 2023MNRAS.526.1646D,2024ApJ...971...90D}.
\par

Neither of these research communities has utilized TI zones, identifiable regions within a phase diagram where the gas is thermally unstable that follow mathematically from the
stability criteria identified by \citet{1986ApJ...303L..79B}.
TI zones have proven very useful for understanding condensation formation within outflows irradiated by an active galactic nucleus \citep{2020ApJ...893L..34D, 2021ApJ...914...62W}.
Recently, they were shown by \citet{2023FrASS..1098135W} to have dynamical importance, as the nonlinear saturation phase of TI begins when gas crosses the boundary of a TI zone.
Position-dependent heating functions can nevertheless make a phase diagram based analysis of TI zones overly complicated, an impediment we seek to remedy in this paper.  
\par
 
By introducing the concept of a critical cooling rate, $\Lambda_c$, this work aims to provide a reworking of Balbus' criteria for TI to arrive at a more convenient and intuitive framework for analyzing cooling functions (with or without position-dependent heating terms and including thermal conduction).  
In \S{2}, we review how to define equilibrium curves and TI zones using a phase diagram-based analysis applied to both ISM and solar cooling functions.
In \S{3}, after deriving $\Lambda_c$, we show that cooling curve plots can be used to perform a stability analysis that is equivalent to that using phase diagrams.  
We further illustrate how thermal conduction can dramatically change the location of TI zones.
In \S{4}, we apply our $\Lambda_c$-based analysis to the coronal loop simulations of \citet{2019ApJ...884...68K} to show that TNE loops are stable to TI modes and yet linearly unstable to catastrophic cooling modes.  
As we discuss in \S{5}, our analysis offers strong support of the claim we made in \citetalias{2025SoPh..300....5W}, which is that the formation of condensations in TNE solutions is triggered by catastrophic cooling instability. 
\par

\section{Cooling functions and their TI zones}
As we alluded to above, there is little emphasis in the literature on the nonequilibrium stability analysis of cooling functions, by which we mean an analysis based on \cite{1986ApJ...303L..79B}'s instability criteria for TI as opposed to the original ones identified by \citet{1965ApJ...142..531F}.
After reviewing commonly adopted functional forms for astrophysical cooling functions in \S{2.1}, we demonstrate how to perform this analysis using TI zones in \S{2.2}.
Then in \S{2.3}, we show how thermal conduction modifies \cite{1986ApJ...303L..79B}'s criteria for TI.  
\par

\subsection {$\Lambda(T)$-based cooling functions}
\label{sec:cool-func}

\begin{figure*}[ht!]
\centering
\plotone{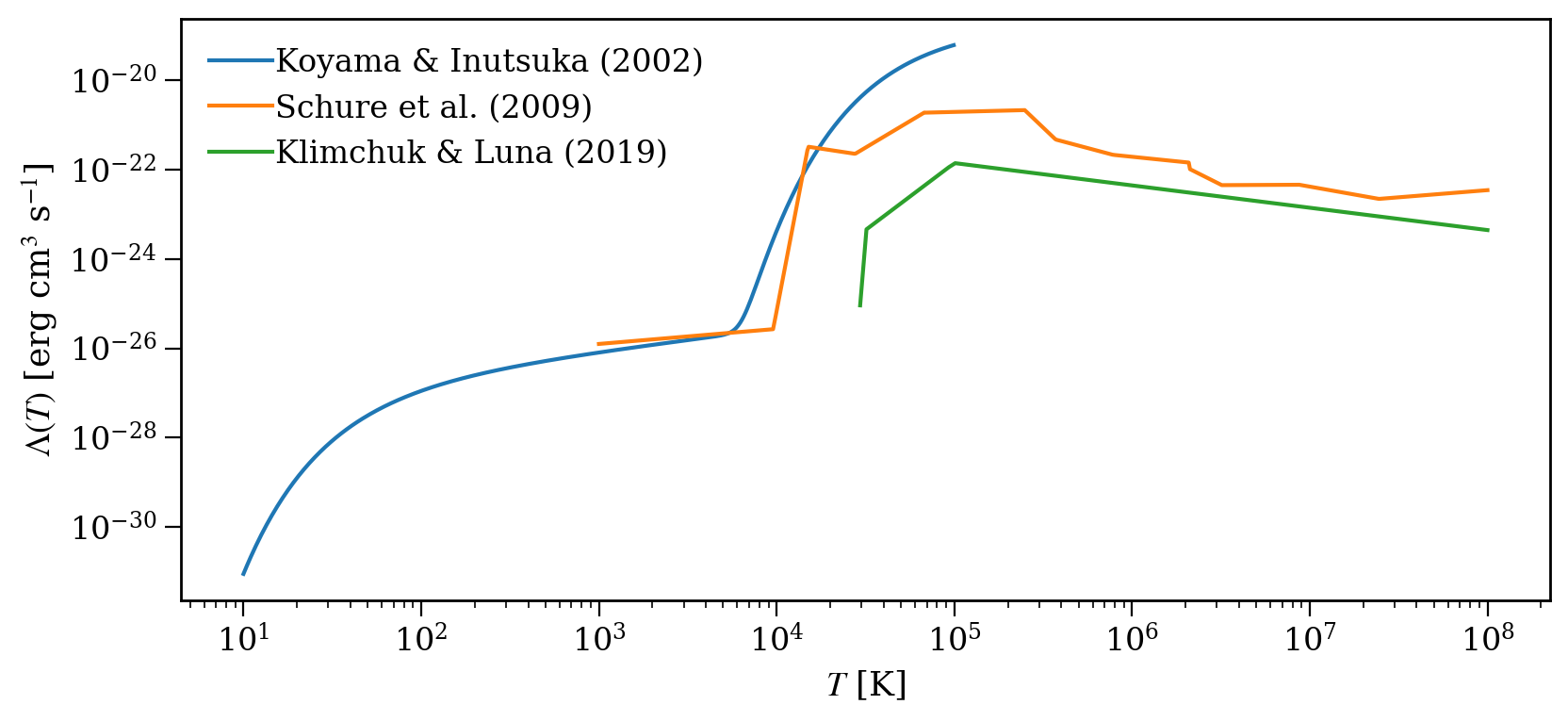}
\caption{Comparison of three different radiative loss functions, $\Lambda(T)$: a commonly used one for the ISM computed by \citet{2002ApJ...564L..97K}, one used to model the solar corona based on collisional ionization equilibrium from \citet{2009A&A...508..751S}, and a simplified curve used for coronal loop modeling by \citet{2019ApJ...884...68K}.
}
\label{fig:compare}
\end{figure*}

The most commonly used prescription for incorporating cooling processes in numerical simulations is to separate out density and temperature dependencies in the form $n^2 \Lambda(T)$, where $n$ is the particle number density and $\Lambda(T)$ is the radiative loss function.
This separation relies on the gas being optically thin to the dominant line-cooling radiation.  
Namely, when there is very weak ionization balance coupling to the radiation field, radiative processes such as photoionization heating and Compton heating and cooling can be neglected.  
The remaining processes, such as line cooling and collisional ionization, depend on two-body interactions, which results in cooling rates with an $n^2$ density dependence \citep{2003adu..book.....D}. 
\par

We apply our stability analysis to two substantially different astrophysical environments: the ISM and the solar corona.  
We adopt the widely-used analytic ISM loss function from \citet{2002ApJ...564L..97K}, and a piecewise-fit to the solar SPEX\_DM tabular loss function presented by \citet{2009A&A...508..751S}.
In addition, the simulation data in \S{4} uses a simplified piecewise loss function from \citet{2019ApJ...884...68K}.  
All three radiative loss functions are compared in Fig.~\ref{fig:compare}.
Note that while the cooling curve from \citet{2002ApJ...564L..97K} spans a significantly lower range of cooling rates compared to the two solar functions, the rates mostly agree with those of \citet{2009A&A...508..751S} in the vicinity of $T \sim 10^4$ K, which corresponds to the temperature at which neutral hydrogen begins to ionize.  
The explicit forms for each of these radiative loss functions is given in Appendix~\ref{app:1}.
\par

Various forms for a heating function have been employed in the literature.
The heating rate dependence on density and temperature is largely unknown in the case of the solar corona due to the uncertainties in the dissipative heating rates of kinetic plasma processes.  
Similarly, the dominant heating mechanisms in the cold and warm ISM phases is difficult to determine because it involves complicated processes like the photoelectric heating of dust grains, cosmic ray ionization, and X-ray photoionization \citep[see][]{1972ARA&A..10..375D, 2020SSRv..216...68G}.
The \cite{2002ApJ...564L..97K} and \citet{2009A&A...508..751S} radiative loss functions, upon being combined with a simple parametrization of the heating rate, as given below, are hereafter respectively referred to as simply KI and SPEX.
\par

The net cooling function, denoted $\mathcal{L} = \mathcal{L}(\rho,T)$, represents the sum of energy losses and gains in units of $\rm{erg\,g^{-1}\,s^{-1}}$.
This enters the internal energy equation for non-adiabatic, ideal MHD as the source term $-\rho\mathcal{L}$: 
\begin{equation}
   \rho\frac{D\mathcal{E}}{Dt} = -p\mathbf{\nabla} \cdot \gv{v} - \rho \mathcal{L} + \nabla \cdot [\kappa(T) \mathbf{\nabla} T].
   \label{eq:energy}
\end{equation}
Here, $\rho$, $\gv{v}$, and $p$ are the gas density, velocity, and pressure, respectively, $\mathcal{E} = c_V T$ is the gas internal energy with $c_V$ the specific heat at constant volume, and $\kappa(T) = \chi T^{5/2}$ is the parallel thermal conductivity, assumed to be the classical Spitzer value with $\chi = 6.8 \times 10^{-7} $erg s$^{-1}$ K$^{-7/2}$ cm$^{-1}$.
\par

The source term for both KI and SPEX can be parameterized as
\begin{equation}
    \rho \mathcal{L} = n^2 \Lambda(T) - C_H \rho^a,
\label{eq:rhoL}
\end{equation}
where $C_H$ and $a$ are constants.
With $\rho = \bar{m} n$, where $\bar{m}$ is the mean particle mass, solving for $\mathcal{L}$ gives
\begin{equation}
    \mathcal{L} = \frac{\rho}{\bar{m}^2} \Lambda(T) - C_H \rho^{a-1}.
\label{eq:CIEfunc_simp}
\end{equation}
It is convenient to define a net cooling function $L_{net}$ having the same units as $\Lambda(T)$, namely $\rm{erg} \; \rm{cm^3} \; \rm{s^{-1}}$:
\begin{equation}
    \label{eq:lnet}
    L_{net} \equiv \frac{\bar{m}}{n}\mathcal{L} = \Lambda(T) - \Gamma(n).
\end{equation}
Here we have introduced a heating function in these units also; comparing with \eqref{eq:CIEfunc_simp}, 
\begin{equation}
    \label{eq:Gamma}
    \Gamma(n) = C_H \bar{m}^a  n^{a-2} .
\end{equation}
\par

The KI radiative loss function given in Appendix~\ref{app:1} is a corrected version given by \citet{2007ApJ...657..870V}; for the heating function, using our notation, these authors set $a = 1$ and $C_H\bar{m} = 2.0 \times 10^{-26}$ erg s$^{-1}$.  
For SPEX, we adopt a common practice of making the heating function dependent on the initial conditions of the plasma \citep[e.g.][]{2015ApJ...812...92N, 2020A&A...639A..20K, 2021A&A...655A..36H, 2021ApJ...920L..15R}.  
Following \citet{2020ApJ...897...64Y}, \citet{2021A&A...655A..36H}, and \citet{2023SoPh..298..102M}, here we set $a=0$; when this choice is adopted for local periodic box simulations that begin in thermal equilibrium at some given $n_0$ and $T_0$, \eqref{eq:Gamma} gives $C_H = n_0^2 \Lambda(T_0)$.  Here we take $T_{0} = 10^6$ K and $n_{0} = 10^9 ~\rm{cm}^{-3}$.
The choice $a=0$ results in $\Gamma(n) \propto n^{-2}$, and as will become clear in \S{4}, this happens to be the scaling used to obtain TNE solutions in global simulations of coronal loops \citep[e.g.][]{2011ApJ...737...27X,2013ApJ...771L..29F,2013ApJ...773...94M, 2015ApJ...807..142F, 2017ApJ...844...26F, 2017ApJ...845...12K, 2018A&A...618A.135R, 2018ApJ...855...52F, 2019A&A...625A.149J, 2022A&A...658A..71P}.
\par

\subsection{TI zones}
This paper builds off the work of \citet{1986ApJ...303L..79B}, who generalized the instability criteria derived by \citet{1965ApJ...142..531F} by considering the effect of perturbations in a homogeneous gas with $\mathcal{L} \neq 0$.  
He showed that the isobaric and isochoric instability criteria are modified to
\begin{align}
\label{eq:isobar}
\left( \frac{\partial \mathcal{L}/T}{\partial T}\right)_p < 0 \;(\textnormal{isobaric}),
\\
\label{eq:isochor}
\left( \frac{\partial \mathcal{L}/T}{\partial T}\right)_\rho < 0 \; (\textnormal{isochoric}).
\end{align}
These criteria reduce to those found by Field, namely $(\pdtext{\mathcal{L}}{T})_p < 0$ and $(\pdtext{\mathcal{L}}{T})_\rho < 0$, when $\mathcal{L} = 0$.
They follow from linear theory by finding the actual growth/damping rates of perturbations, and our stability analysis is aided by returning to these rates. 
First we define the thermal time as
\begin{equation}
    t_{\rm th} \equiv \frac{c_V T}{n\Em/\bar{m}}.
\end{equation}
Here we have followed \citet{1990ApJ...358..375B} in defining the maximum of the local (nonequilibrium) heating or cooling rate as $\Em \equiv \textnormal{max}(\Lambda(T), \Gamma(n))$.
Then with this definition, the solution to the dispersion relation governing TI modes will yield growth rates on the order of $t_{\rm th}^{-1}$ times one of the following dimensionless quantities \citep[see][]{2019ApJ...875..158W}
\begin{align}
    N_p' &= \frac{T}{n\Em/\bar{m}} \left[ T\left(\frac{\partial\mathcal{L}/T}{\partial T}\right)_p \right]; \\
    N_\rho' &= \frac{T}{n\Em/\bar{m}}\left[T\left(\frac{\partial\mathcal{L}/T}{\partial T}\right)_\rho\right]. 
\end{align}
In terms of these quantities, Balbus' isobaric instability criterion is $N_p' < 0$, while his isochoric one is $N_\rho' < 0$. 
It should be noted that while \citet{2019ApJ...875..158W} did not use $\Em$ to normalize $N_p'$ and $N_\rho'$ as we do here, the growth/damping rates $N_p'/t_{\rm th}$ and $N_\rho'/t_{\rm th}$ have no actual dependence on $\Em$, and the reason for this choice will become clear in \S{3.1}.
\par

\begin{figure*}[ht!]
\centering
\includegraphics[width=0.9\textwidth]{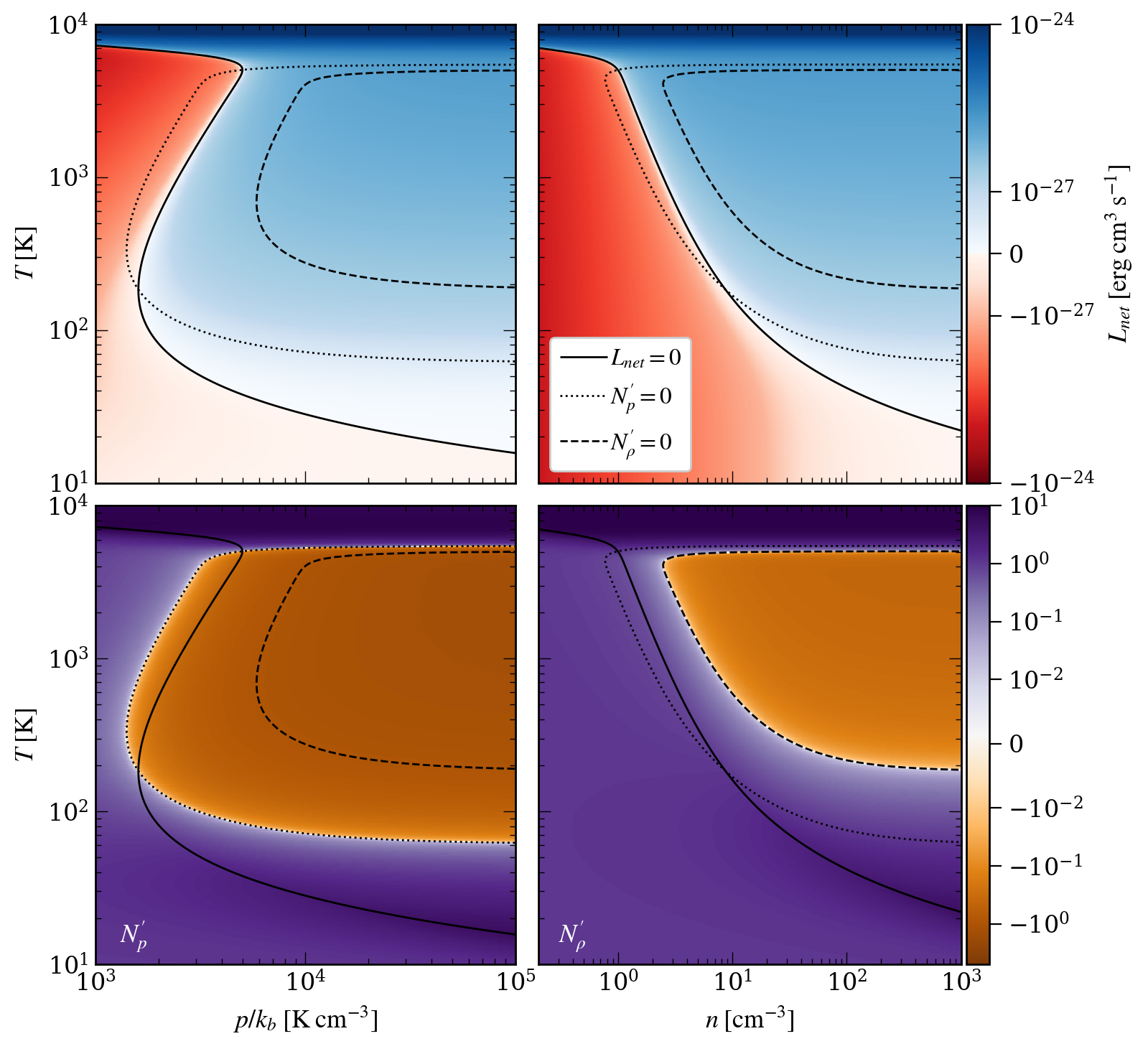}
\caption{TI zones plotted on phase diagrams for the KI cooling function.
\textit{Top panels:} $T$-$p$ and $T$-$n$ phase diagrams, displaying a colormap of the net cooling rate, $L_{net}$, where red indicates heating ($L_{net} < 0$) and blue cooling ($L_{net} > 0$).
The solid, dotted, and dashed lines are the equilibrium curve ($L_{net} = 0$), the isobaric Balbus contour ($N_p' = 0$), and the isochoric Balbus contour ($N_\rho' = 0$), respectively.
\textit{Bottom panels:} Colormaps of $N_p'$ and $N_\rho'$, which are the dimensionless isobaric and isochoric growth rates (orange colors) and damping rates (purple colors).
In $T$-$p$ phase diagrams, positively sloped regions of an equilibrium curve are \textit{isobarically} unstable, while in $T$-$n$ phase diagrams, positively sloped regions are \textit{isochorically} unstable.
Note that while the slope of the equilibrium curves are entirely negative in the right panels, there is nevertheless a prominent region of isochoric instability above the equilibrium curve (the orange region in the lower right panel, which defines an isochoric TI zone).}
\label{fig:koy_phase}
\end{figure*}

We cast these expressions in terms of $L_{net}$ by first expanding $[\partial(\mathcal{L}/T)/\partial T]_p =(1/T)(\partial \mathcal{L}/\partial T)_p -\mathcal{L}/T^2$.  Then noting the identity 
\begin{equation}
    \label{eq:th_ident}
    \left( \frac{\partial \mathcal{L}}{\partial T}\right)_p  = \left( \frac{\partial \mathcal{L}}{\partial T}\right)_\rho - \frac{\rho}{T}\left( \frac{\partial \mathcal{L}}{\partial \rho}\right)_T
\end{equation}
and using \eqref{eq:lnet}, we have
\begin{align}
    \begin{split}
    N_p' =& \frac{1}{\Em} \bigg[T\left(\frac{\partial L_{net}}{\partial T}\right)_n - n\left(\frac{\partial L_{net}}{\partial n}\right)_T - 2L_{net}\bigg]; 
    \label{eq:Np_dim}
    \end{split}
    \\
    \begin{split}
    N_\rho' =& \frac{1}{\Em}\bigg[T\left(\frac{\partial L_{net}}{\partial T}\right)_n - L_{net}\bigg]. 
    \label{eq:Nrho_dim}
    \end{split}
\end{align}
The contours $N_p' = 0$ and $N_\rho' = 0$ define the boundaries of TI zones (on which cooling rates reach their critical values, as will be shown in \S{3.1}).
The significance of these boundaries is this: only within the isochoric TI zone is Balbus' criterion for isochoric instability satisfied, and similarly for the isobaric TI zone \citep{2023FrASS..1098135W}.
In Fig.~\ref{fig:koy_phase} and~\ref{fig:spex_phase}, we map these zones (using the numerical methods described in Appendix~\ref{app:2}), thereby showing instability locations relative to equilibrium curves (the contours $L_{net} = 0$).
\par 

TI zones are displayed in two different ways in Fig.~\ref{fig:koy_phase}.  
The top panels compare the zero contours of $N_p' = 0$ and $N_\rho' = 0$ plotted on colormaps of $L_{net}$ and displayed on both temperature-pressure ($T$-$p$) and temperature-density ($T$-$n$) phase diagrams.
The bottom panels show colormaps of $N_p'$ and $N_\rho'$ to reveal the actual structure of TI zones. 
Orange (purple) colors denote unstable (stable) regions where $N_p'$ and $ N_\rho'$ are perturbation growth (damping) rates.
Notably, unstable parameter space is mostly confined to regions with positive $L_{net}$ (i.e., where gas is radiatively cooling).
\par 

\begin{figure*}[ht!]
\centering
\includegraphics[width=0.9\textwidth]{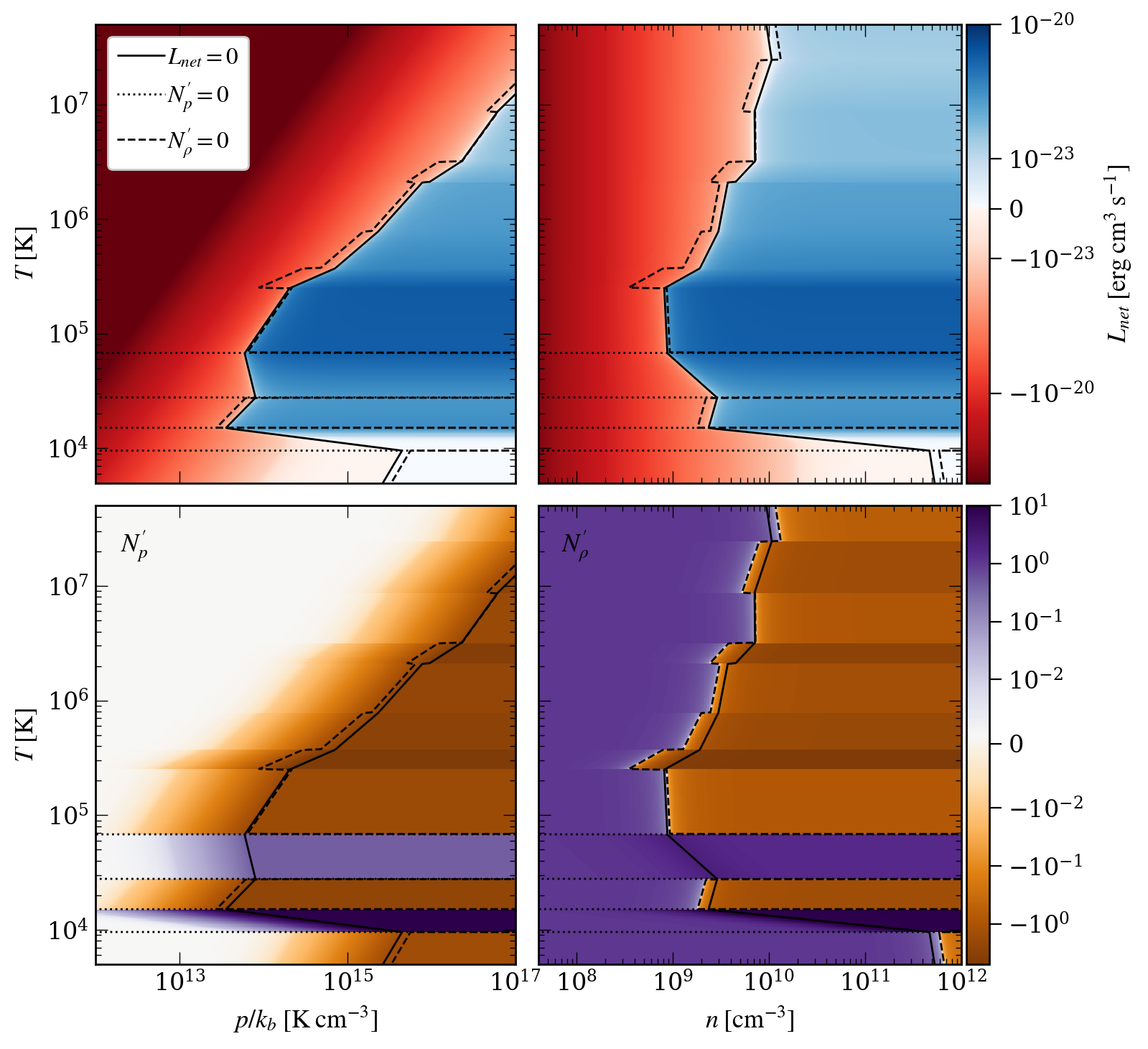}
\caption{Same as Fig.~\ref{fig:koy_phase} but for the SPEX cooling function.
Unlike in Fig.~\ref{fig:koy_phase}, the equilibrium curve has positive sloping regions in the $T$-$n$ phase diagram, indicating that both the isobaric and isochoric instability criteria are satisfied.  Such a slope-based analysis tells one nothing about the stability of regions off the equilibrium curve.  By plotting TI zones, we recover the sloped-based results but also see where there is instability in nonequilibrium parameter space.}
\label{fig:spex_phase}
\end{figure*}

Fig.~\ref{fig:spex_phase} shows how coronal heating rates result in vastly different stability properties for the net cooling functions that are commonly adopted. 
Notice in particular that SPEX features an equilibrium curve that lies almost entirely within the isobaric and isochoric TI zones. 
The isobaric TI zone happens to be density-independent and thus appears as a series of horizontal lines spanning the diagram. 
This occurrence is due to the nature of the SPEX heating function, which again comes from \eqref{eq:Gamma} with $a = 0$:
\begin{equation}
    \Gamma(n) = \Lambda(T_0) \left(\frac{n_0}{n}\right)^{2}.
\end{equation}
The derivative of the heating function is
\begin{equation}
    \label{eq:heat_deriv}
    \frac{d\Gamma(n)}{dn} = \frac{-2\Gamma(n)}{n}.
\end{equation}
Upon plugging this into \eqref{eq:Np_dim}, the dimensionless growth rate $N_p^\prime$ loses its density dependence.
\par

The caption to Fig.~\ref{fig:koy_phase} explains how to do a customary stability analysis of locations along the equilibrium curve based on the slope of that curve.  
From this analysis one would conclude that KI is everywhere isochorically stable while SPEX is subject to isochoric instability at points along the equilibrium curve that have a positive slope in the $T$-$n$ plane.  
In actuality, both the ISM and coronal cooling functions exhibit extensive isochoric TI zones away from equilibrium.  
The customary slope-based analysis fails to convey this information.   
Dynamical processes such as shock heating or magnetic reconnection can easily push plasma off the equilibrium curve and into these unstable regions, hence why it is important to compute TI zones.  
In \S{3.2}, we furthermore show how to analyze specified evolutionary paths through this unstable parameter space.   
\par

\subsection{Instability criteria with thermal conduction}
The locations of the TI zones calculated above can be quite sensitive to the thermal conductivity of the gas (see \S{3.3}).
Thermal conduction along magnetic field lines will have a strongly stabilizing effect on all TI modes with wavelengths smaller than the Field length, 
\begin{equation}
    \label{eq:field_l}
    \lambda_F = 2\pi\sqrt{\frac{\kappa(T) T}{n^2\Em}}.
\end{equation}
The Field length represents the scale at which a heat flux leads to evaporation overcoming condensation, thereby damping perturbations below this scale \citep{1990ApJ...358..375B}.
Setting $T_6 = T/10^6$ and $n_9 = n/10^9$, we can evaluate the Field length for typical conditions in a coronal loop:
\begin{equation}
    \label{eq:field_l2}
    \lambda_F = 5.2 \times 10^{10} \frac{T_6^{7/4}}{n_9\sqrt{\Em/10^{-23}}} \;\;\;\; [\rm{cm}] .
\end{equation}
This value is much larger than the average loop length of $\sim 10^9 - 10^{10} \rm{cm}$.  
Thus, thermal conduction is expected to stabilize TI modes for all but the longest loops.
\par

\citet{2019ApJ...875..158W} showed that thermal conduction modifies Balbus' instability criteria to the stricter conditions 
\begin{equation}
    \label{eq:Np}
    N_p' < -\left(\frac{\lambda_F}{\lambda}\right)^2; 
\end{equation}
\begin{equation}
    \label{eq:Nrho}
    N_\rho' < -\left(\frac{\lambda_F}{\lambda}\right)^2.
\end{equation}
Clearly, we recover Balbus' criteria for both $\lambda \gg \lambda_F$ and $\lambda_F \rightarrow 0$.  
Since the dimensionless values $N_p'$ and $N_\rho'$ are order unity or smaller quantities according the bottom panels of Fig.~\ref{fig:spex_phase}, we see mathematically how thermal conduction will suppress TI when $\lambda \lesssim \lambda_F$.  
It is therefore necessary to account for how the stabilizing effect of thermal conduction changes the location of TI zones (see \S{3.3}), the boundaries of which are now the wavelength-dependent contours $N_p^\prime = -(\lambda_F/\lambda)^2$ and $N_\rho^\prime = -(\lambda_F/\lambda)^2$.
\par

\section{Results}
\label{sec:results}
In \S{2}, we demonstrated how to calculate TI zones, which facilitate the stability analysis of cooling functions in nonequilibrium regions of parameter space where $L_{net} \neq 0$.  
In this section, we introduce the concept of a critical cooling rate, denoted $\Lambda_c$, that permits an equivalent but more intuitive stability analysis.
Here we limit our analysis to the $\Lambda(T)$-based cooling functions defined in \S{2.1}, but we account for arbitrary temperature and density dependence, i.e. for $L_{\rm net} = \Lambda(T,n) - \Gamma(T,n)$, in Appendix~\ref{app:2}.
\par

\subsection{Critical cooling rates}
By substituting \eqref{eq:Np_dim} and \eqref{eq:Nrho_dim} into the instability criteria in \eqref{eq:Np} and \eqref{eq:Nrho}, and then eliminating $L_{net}$ using \eqref{eq:lnet}, we find
\begin{align}
    \begin{split}
        \label{eq:isobar_TC_ineq}
         \frac{d\Lambda(T)}{dT} &+ \frac{n}{T}\frac{d\Gamma(n)}{d n} - \frac{2\Lambda(T)}{T} + \frac{2\Gamma(n)}{T} \\
        &< -\frac{\Em}{T}\left(\frac{\lambda_F}{\lambda}\right)^2;
    \end{split}
\end{align}
\begin{align}
    \begin{split}
    \label{eq:isochor_TC_ineq}
     \frac{d\Lambda(T)}{dT} &- \frac{\Lambda(T)}{T} + \frac{\Gamma(n)}{T} < - \frac{\Em}{T}\left(\frac{\lambda_F}{\lambda}\right)^2.
    \end{split}
\end{align}
Solving for $\Lambda(T)$ gives
\begin{align}
    \begin{split}
        \label{eq:isobar_ineq}
        \Lambda(T) &> \frac{1}{2} \bigg[T\frac{d\Lambda(T)}{dT}  + n \frac{d\Gamma(n)}{d n} + 2\Gamma(n)  \\ &+ \Em\left(\frac{\lambda_F}{\lambda}\right)^2\bigg];
    \end{split}
    \\\notag
    \\ 
    \begin{split}
        \label{eq:isochor_ineq}
        \Lambda(T) &> T\frac{d\Lambda(T)}{dT} + \Gamma(n) + \Em\left(\frac{\lambda_F}{\lambda}\right)^2. 
    \end{split}
\end{align}
\begin{figure*}[ht!]
    \centering
    \plotone{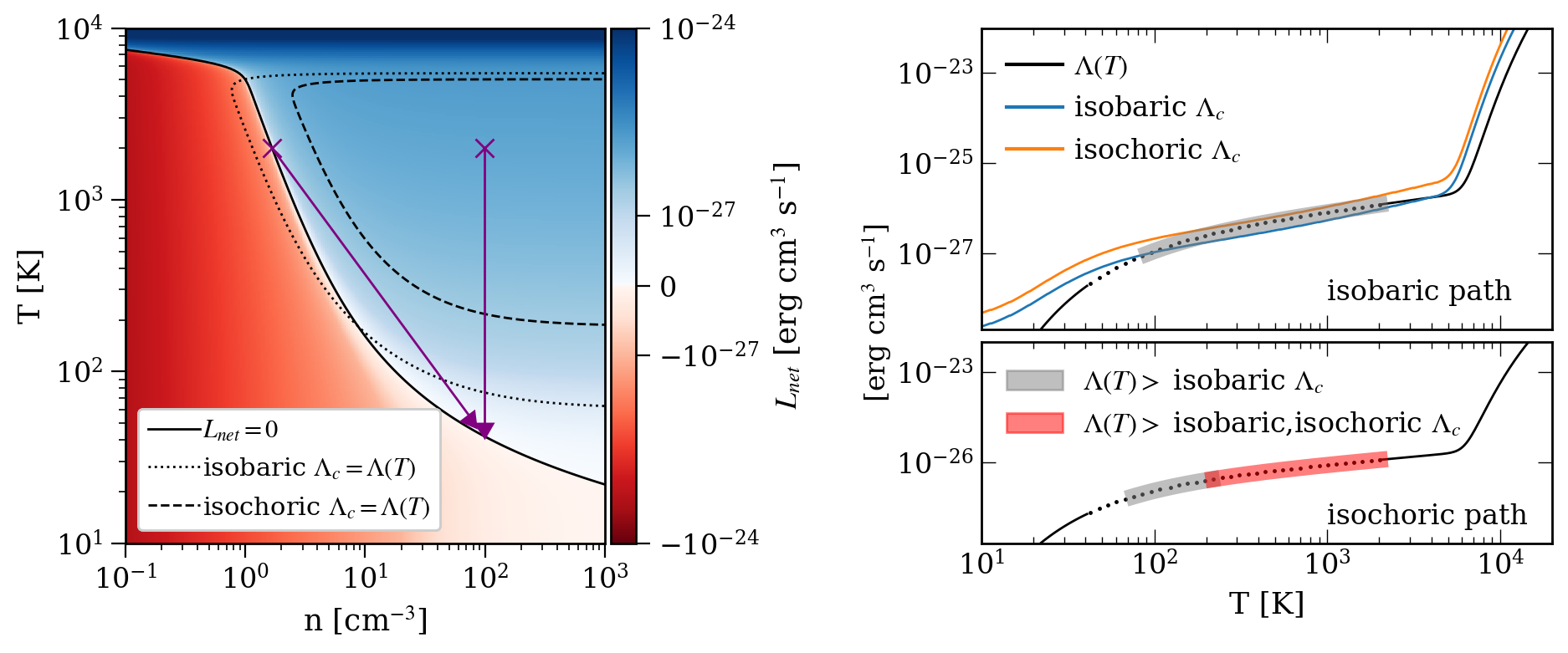}
    \caption{$\Lambda_c$-based nonequilibrium analysis of the KI cooling function.
    \textit{Left panel:} An equivalent $T$-$n$ phase diagram to the top right panel of Fig.~\ref{fig:koy_phase}, this time also showing two nonequilibrium evolutionary paths: (i) an isobaric path from an unstable state to a stable state (diagonal arrow); and (ii) an isochoric path from a hot state back to an equilibrium state.  
    \textit{Right panels:} The $\Lambda(T)$ vs. $T$ curve from Figure~\ref{fig:compare} is reproduced here as the solid black lines.
    The dotted portions of these lines correspond to the temperatures spanned by the isobaric and isochoric paths from the left panel. 
    We arrive at the highlighted portions by first plotting the blue and orange lines, which represent the isobaric and isochoric $\Lambda_c$ values (\eqref{eq:isobar_L} and~\eqref{eq:isochor_L}, respectively). 
    Gray (red) highlights denote when $\Lambda(T) > \Lambda_c$ for the isobaric value (both values) of $\Lambda_c$.  Comparing the highlighted regions with the purple paths in the left panel, notice how the isobaric path avoids passing through the isochoric TI zone, hence it is only shaded gray, while the isochoric path starts off in the isochoric TI zone and exits this zone when the highlighting changes color from red to gray.  From there, it changes from isobarically unstable to stable (unshaded) upon exiting the isobaric TI zone.}
    \label{fig:koy_Lc}
\end{figure*}Thus, upon defining the right hand side expressions as critical cooling rates, we arrive at the simple instability criterion
\begin{align}
    \label{eq:lc_eq.}
    \Lambda(T) &> \Lambda_c,
\end{align}
where the critical rates are given by
\begin{align}
    \begin{split}
    \label{eq:isobar_L}
        (\rm{isobaric}) \; \Lambda_c &= \Gamma(n) + \frac{1}{2} \bigg[T\frac{d\Lambda(T)}{dT}  + n \frac{d\Gamma(n)}{d n}  \\ &+ \Em\left(\frac{\lambda_F}{\lambda}\right)^2\bigg];
    \end{split}
    \\
    \begin{split}
    \label{eq:isochor_L}
        (\rm{isochoric}) \; \Lambda_c &= \Gamma(n) + T\frac{d\Lambda(T)}{dT} + \Em\left(\frac{\lambda_F}{\lambda}\right)^2.
    \end{split}    
\end{align}
The above instability criterion is an equivalent alternative to \eqref{eq:Np} and \eqref{eq:Nrho}.
Separate critical cooling rates now replace the traditional isobaric and isochoric inequalities, and TI zone boundaries become the contours $\Lambda(T) = \Lambda_c$ rather than 
$N_p^\prime = -\left(\lambda_F/\lambda\right)^2$ and 
$N_\rho^\prime = -\left(\lambda_F/\lambda\right)^2$.
\par

Notice that had we chosen $\Lambda(T)$ instead of $\Em$ as a normalization factor in \eqref{eq:Np_dim} and \eqref{eq:Nrho_dim}, the thermal conduction term would formally modify the simple inequality $\Lambda(T) > \Lambda_c$ into something more complicated.  This would be incorrect, however, since $\Em$ enters into the definition of $\lambda_F$ also and therefore $\Lambda_c$ does not depend on $\Em$. 
\par

When considering \eqref{eq:isobar_L} and \eqref{eq:isochor_L}, if a term leads to an increase in the value of $\Lambda_c$, it has a stabilizing influence on the gas.
Notice the (de)stabilizing influence of the derivative $d\Lambda(T)/dT$ is half as great under isobaric conditions compared to isochoric ones.
In particular, from Fig.~\ref{fig:compare} we see that the KI cooling rate derivative is positive, thereby making the isochoric TI zone more difficult to enter than the isobaric one.  In contrast, a significant portion of the SPEX cooling curve has a negative slope.
The heating prescription adopted by the many authors listed at the end of \S{2.1} also introduces a negative heating rate derivative (see~\eqref{eq:heat_deriv}), which has a major destabilizing effect for the isochoric value of $\Lambda_c$.
\par

\begin{figure*}[t!]
\centering
\plotone{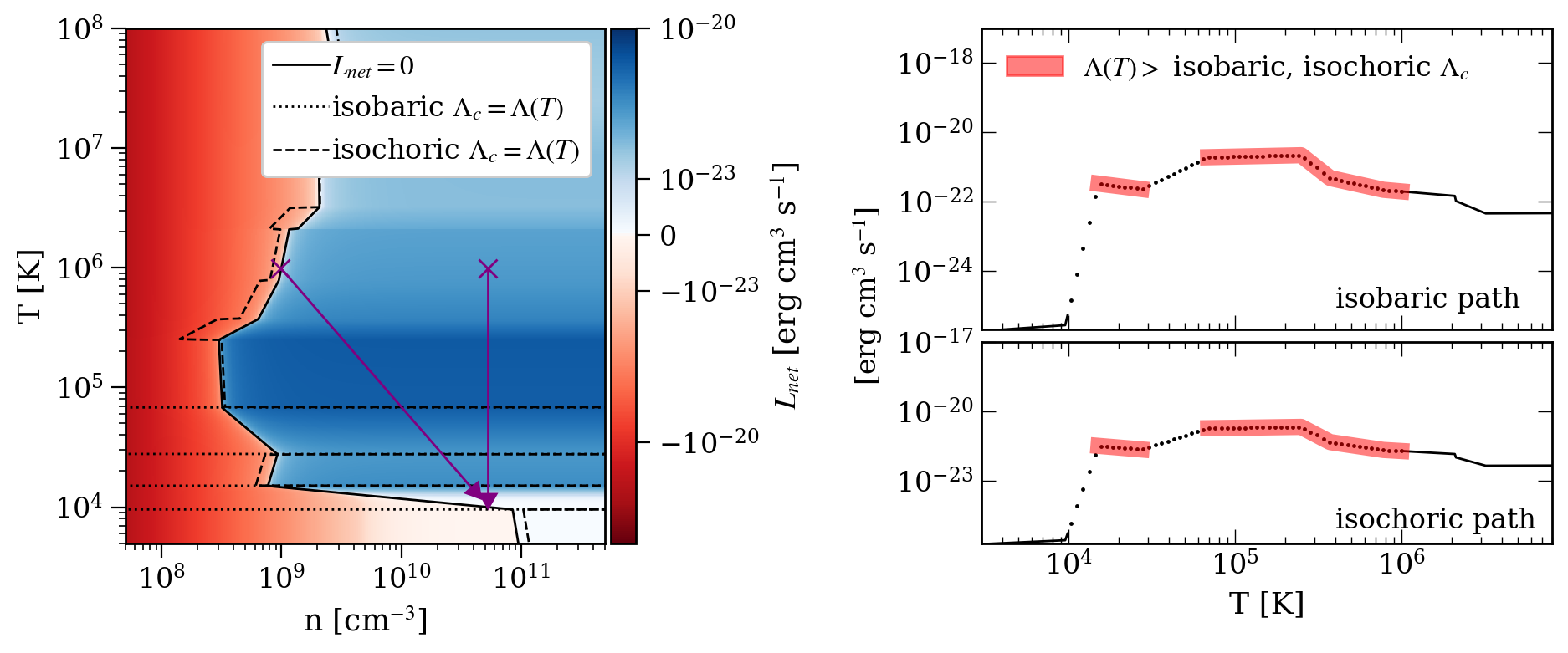}
\caption{Same as Figure~\ref{fig:koy_Lc}, showing the SPEX cooling function.
For simplicity, here we omit the blue and orange curves in the top-right panel.  Again, red highlights indicate TI zones where $\Lambda(T) > \Lambda_c$ for both the isobaric and isochoric values of $\Lambda_c$. 
}
\label{fig:spex_Lc}
\end{figure*}

Using this new $\Lambda_c$-based formalism, the left panels of Fig.~\ref{fig:koy_Lc} and Fig.~\ref{fig:spex_Lc} present an equivalent stability analysis to that given earlier in the top right panels of Fig.~\ref{fig:koy_phase} and Fig.~\ref{fig:spex_phase}.  
A comparison with the right panels in Fig.~\ref{fig:koy_Lc} and Fig.~\ref{fig:spex_Lc} illustrates how a phase diagram based analysis can be replaced with an equivalent cooling curve plot, as we now detail.
\par

\begin{figure*}[ht!]
\centering
\includegraphics[width=0.82\textwidth]{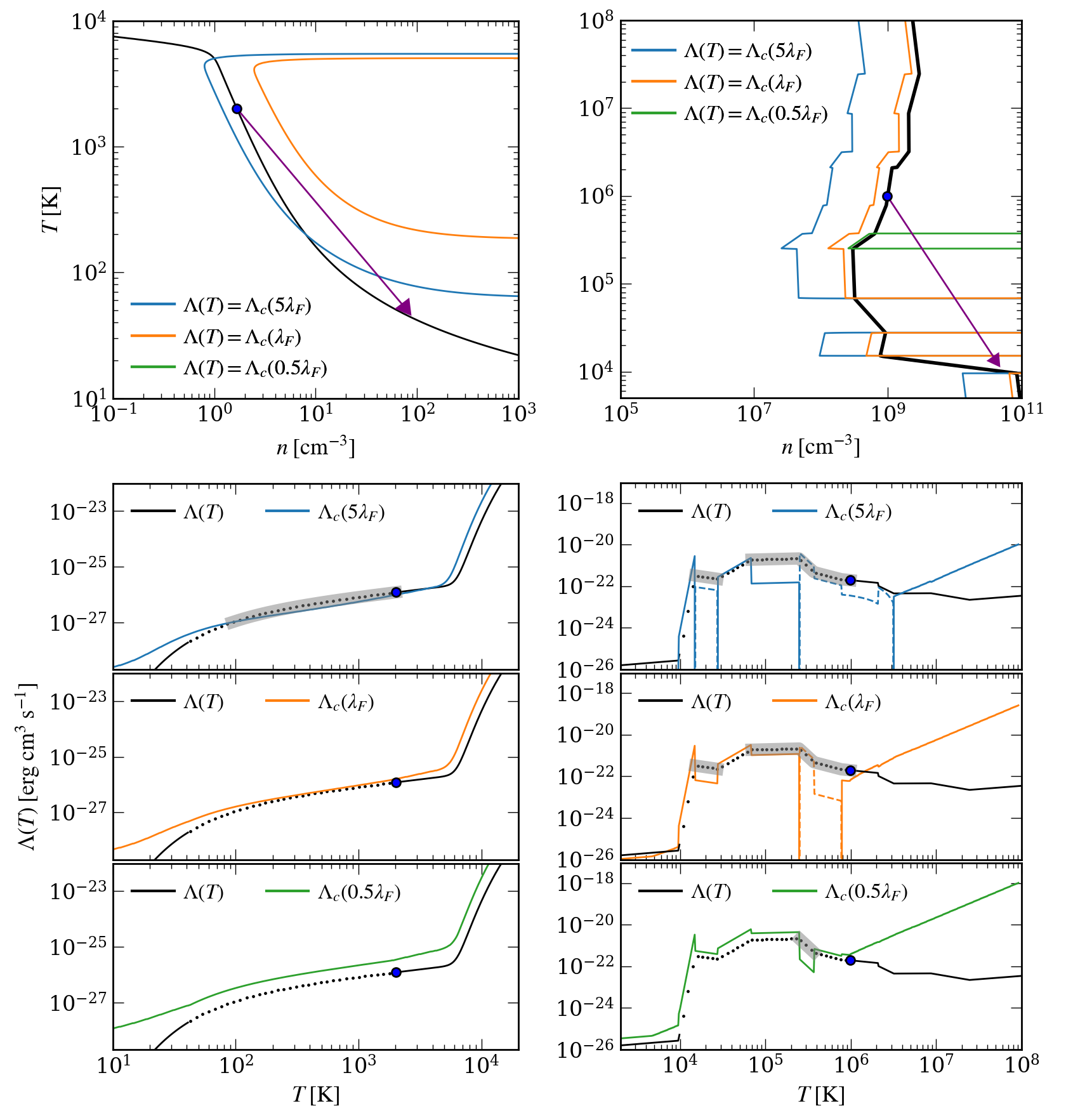}
\caption{Analysis of the stabilizing effect of thermal conduction on isobaric TI zones for the KI and SPEX cooling functions.
\textit{Top panels:} $T-n$ phase diagrams showing equilibrium curves (solid black lines)
and the isobaric contours $\Lambda(T) = \Lambda_c(5\lambda_F)$, $\Lambda_c(\lambda_F)$, and $\Lambda_c(0.5\lambda_F)$ (blue, orange, and green lines, respectively).
Notice how the isobaric TI zone grows smaller with increasing thermal conduction (the green contour in the KI phase diagram falls off of the plot completely).   
The purple arrow shows the same isobaric paths from Figure~\ref{fig:koy_Lc} and Figure~\ref{fig:spex_Lc}.
TI zones for SPEX are no longer horizontal lines and in fact resemble the \textit{isochoric} TI zones in Figure~\ref{fig:spex_Lc} because $\lambda_F$ introduces density dependence.
\textit{Bottom panels:} Cooling curve plots comparing $\Lambda(T)$ and isobaric $\Lambda_c$ for each value of $\lambda$ used to evaluate the TI zones in the top panel. The strength of thermal conduction increases from top to bottom, illustrating stabilization for $\lambda \lesssim \lambda_F$. Gray highlights have the same meaning as in Figure~\ref{fig:koy_Lc}.
}
\label{fig:TC}
\end{figure*}

\subsection{Nonequilibrium stability analysis}
Once again, Field's instability criteria apply only to gas in an equilibrium state with heating and cooling in balance ($L_{net} = 0$).  
In numerical simulations, even gas that has reached a steady state will have $L_{net} \neq 0$, including completely static solutions (provided there is a nonzero heat flux\textemdash see \eqref{eq:TCbalance}).  
Despite this fact, Field's criteria are still regularly employed to analyze the stability of dynamically evolving gas \citep{2015ApJ...805...73I, 2016MNRAS.457.2554C, 2020A&A...636A.112C, 2023SoPh..298..102M, 2024ApJ...976..226S, 2024PhRvE.110f5201F, 2025ApJ...981..190S}.
Here we demonstrate how to do a dynamical stability analysis when $L_{net} \neq 0$ for two distinct scenarios: 
(i) isobaric evolution away from one equilibrium state toward another, and 
(ii) isochoric evolution following a rapid heating event that displaces gas far above its equilibrium state, from which it proceeds to cool back to equilibrium.
\par

In Fig.~\ref{fig:koy_Lc}, these scenarios are depicted by the purple arrows in the left panel, with the diagonal arrow showing an isobaric path and the vertical arrow an isochoric one, as well as by the the dotted portions of the $\Lambda(T)$ curves in the right panels.  
Both paths begin and end at the same temperature, but due to the difference in density, the gas will have different cooling and heating rates (and hence perturbations will have different damping/growth rates depending on their stability). 
The shaded regions in the right panels indicate transitions through TI zones; see the figure caption for details.
The top right panel displays the isobaric and isochoric values of $\Lambda_c$ as the blue and orange curves, providing a visual depiction of how these TI zones were identified. 
Notice that because $\Lambda(T)$ remains below the isochoric value of $\Lambda_c$ throughout, it must be isochorically stable; this is geometrically consistent with this path avoiding the isochoric TI zone in the left panel. 
\par

From the left panel of Fig.~\ref{fig:koy_Lc}, note that the isochoric TI zone is fully embedded within the isobaric TI zone, which implies that gas is isobarically unstable whenever it is isochorically unstable.  
This is reflected in the bottom right panel: the red portion of the path satisfies $\Lambda(T) >\Lambda_c$ for both values of $\Lambda_c$.  
\par

Fig.~\ref{fig:spex_Lc} presents the same analysis for the SPEX function. 
In this case, both paths occupy the isochoric TI zone from the start, pass through the stable region between the two horizontal lines (the purple region in Fig.~\ref{fig:spex_phase}), and then again enter a TI zone before reaching a stable branch of the equilibrium curve.  
Accordingly, they have equivalent shadings in the panels at right.
\par

\subsection{Stabilization of TI modes due to\\thermal conduction}
As discussed in \S{2.3}, the thermal conduction term completely stabilizes TI for wavelengths smaller than the Field length.
This term makes $\Lambda_c$ a function of $\lambda$ and thus the locations of TI zones are wavelength-dependent.  
Confining our attention to isobaric TI zones, 
Fig.~\ref{fig:TC} illustrates the effect of thermal conduction on TI zones for both KI and SPEX.
The top panels show the equilibrium curve (black lines) plotted on a $T$-$n$ phase diagram along with the isobaric path studied in \S{3.2} (purple arrow).
\par

Focusing on the top left panel, the blue, orange, and green curves are three different TI zones defined by the contours
$\Lambda(T) = \Lambda_c(5\lambda_F)$, 
$\Lambda(T) = \Lambda_c(\lambda_F)$, and
$\Lambda(T) = \Lambda_c(0.5\lambda_F)$.
The blue contour is almost identical to the TI zone without thermal conduction shown in previous figures.  
The isobaric path completely avoids the TI zones outlined by the orange curve, clearly illustrating stabilization due to thermal conduction for $\lambda \lesssim \lambda_F$. 
\par

The bottom left panels in Fig.~\ref{fig:TC} present an equivalent analysis without using a phase diagram, one that would be easier to perform using data from an MHD simulation.
The blue dots mark the starting location of the isobaric path on the equilibrium curve where we evaluated the quoted value of $\lambda/\lambda_F$.
Notice that as the wavelength approaches $\lambda_F$, the location of the dot changes from being unstable to stable according to the instability criterion $\Lambda(T) > \Lambda_c(\lambda)$.  
\par

The right panels of Fig.~\ref{fig:TC} presents the same analysis for SPEX.
The main thing to point out is that because $\lambda_F \propto T$, $\lambda_F/\lambda$ decreases at smaller temperatures along each isobaric path.
This explains why a portion of the path for the curve marked $\Lambda_c(0.5\lambda_F)$ is still unstable.  
Of course, because the location of the blue point is stable for $\lambda = 0.5\lambda_F$, this path would not be taken by a linear perturbation.  
We also note that the isobaric TI zones take on the shape of the isochoric TI zones shown earlier; this is due to the density-dependence of $\lambda_F$.
\par

To decide if a localized region of a given flow field is stable or unstable, what are the longest wavelength TI modes that need to be considered?
In environments like coronal loops, where there is a definite system size, we can safely conclude that any perturbations that can fit in the system will be stable to TI if the Field length exceeds the system size.  
We already employed this argument in \S{2.3} to show that most coronal loops should not be susceptible to TI.  
The textbook answer for general flow fields is that only wavelengths as long as the shortest characteristic gradient length scale among the basic flow variables ($\rho$, $\mathbf{v}$, $p$, or $T$) are relevant, as longer wavelengths will, for example, either be deformed by the velocity gradient or damped by the temperature gradient.
Hence, critical cooling rates evaluated at wavelengths equal to the shortest gradient length scale can serve as a meaningful value to use for computing TI zones in MHD simulations, which we explore further in \S{4.3}.
\par

Though \S{4} focuses on coronal loops, we emphasize that the results in this section can be applied generally to all astrophysical cooling functions, whether analytic or tabulated.
\par

\begin{table*}[th!]
\centering
\movetableright=-1in
\caption {Coronal loop simulations: heating inputs and stability analysis}

\begin{tabular}{cccccc|cccc}
    \hline\hline
    Loop  &$Q_{f}, $ & $Q_{bkg}$ & $\lambda_f$ & $s_{f}$   & $L$  & \multicolumn{2}{c}{Linearly unstable?} & TNE?
        \\ 
       & [erg cm $^{-3}$ s $^{-1}$]  & [erg cm $^{-3}$ s $^{-1}$] &[cm] & [cm] & [cm] & TI modes & CC modes &  &{}\\
    \hline
    
    Loop~1 & $3 \times 10^{-3}$ & $0$ & $2 \times 10^{9}$ & $5 \times 10^{9}$ & $4 \times 10^{9}$ & No & No & No\\
    Loop~2 & $2.66 \times 10^{-3}$ & $2.4 \times 10^{-4}$ & $8 \times 10^{8}$ & $4.5 \times 10^{9}$ & $4\times 10^{9}$ & No & Yes & Yes\\
    \hline
    \hline
\end{tabular}
\label{table1}
\end{table*} 

\section{Application to Coronal Loops}

We have thus far only considered localized heating rates in which $\Gamma = \Gamma(n)$ is position-independent.  
Triggering coronal rain formation using the TNE setup described below requires a spatially dependent heating prescription.  
It is then rather inconvenient to perform a phase diagram based stability analysis, as the equilibrium curve changes at each point along the loop.  
A $\Lambda_c$-based analysis, on the other hand, can be readily applied to this situation.  As it furthermore enables visualizing the stabilizing effect of thermal conduction, it is a useful tool when applied to coronal loop simulations.
\par

\subsection{Klimchuk \& Luna Simulations}
The global 1D coronal loop simulations analyzed below come from \citet{2019ApJ...884...68K} (hereafter referred to as \citetalias{2019ApJ...884...68K}).  They present runs with different types of footpoint heating that yield static, steady-state flow, and TNE solutions (described in \S{4.2}).  
\citetalias{2019ApJ...884...68K} utilized the Adaptively Refined Godunov Solver (ARGOS) code \citep{1999ApJ...512..985A} to conduct 1D hydrodynamic simulations of loops with coronal half-lengths of approximately $L = 40 ~\rm{Mm}$, supplemented by an additional $45 ~\rm{Mm}$ of chromosphere in order to use closed boundaries and at the same time have a realistic dynamical and energetic coupling between the corona and chromosphere. 
The ARGOS code employs an adaptively refined mesh, which adds resolution to follow the onset of condensation formation. 
\par

The simulations in \citetalias{2019ApJ...884...68K} evolve into a variety of thermal states depending on the spatial distribution of heating along the loop.
Each simulation is initialized with an approximate static equilibrium solution under uniform heating, which relaxes to a proper equilibrium before the uniform heating is gradually transitioned to a spatially dependent, exponential heating at $t = 10^5$ s. 
This term allows for differential heating at the footpoints, thereby permitting asymmetric energy deposition.
\par

For the purposes of this study, we restrict our analysis to symmetric solutions and therefore neglect any asymmetry in the heating profile.
Rewriting their Eq. (39) in our notation under the assumption of symmetric heating yields
\begin{align}
    \begin{split}
    \Gamma(n,s) &= \frac{1}{n^2} \bigg[Q_{f}\exp\left(\frac{-(s-s_{f})}{\lambda_f}\right) \\
     &+ Q_{f}\exp\left(\frac{-(s_{f}+2L-s)}{\lambda_f}\right) + Q_{bkg}\bigg],
    \end{split}
    \label{eq:tne_heat}
\end{align}
where $Q_{f}$ is the heating rate for the footpoints, $s$ is the loop position, $s_{f}$ is the initial footpoint position at the top of the chromosphere, $L$ is the chromosphere to loop top half-length, $\lambda_f$ is the heating scale height, and $Q_{bkg}$ is the background heating rate.
The radiative cooling is modeled with a simplified, three-step piecewise radiative loss function $\Lambda(T)$ (reproduced in our notation in Appendix~\ref{app:1}).
\par

We analyzed two simulations performed by Klimchuk and Luna, though not identical to their published runs. 
The static loop solution, hereafter referred to as Loop~1, was subjected to slightly higher footpoint heating ($Q_{f}$) and no background heating ($Q_{bkg}$). 
The TNE solution, hereafter referred to as Loop~2, has a small background heating but a scale height ($\lambda_f$) much smaller than in  Loop~1, so the footpoint-to-apex heating ratio exceeds the threshold for TNE \citepalias{2019ApJ...884...68K}, which is not the case for Loop~1.  
These particular choices are not special. 
The only important factor is the ratio of footpoint-to-apex heating. 
A static solution can be achieved with uniform heating, but we are interested in comparing stratified heating that is above and below the threshold for TNE. 
The slight differences in $s_f$ were a convenience and are not significant.
The input values for the heating function, \eqref{eq:tne_heat}, are listed in Table~\ref{table1}.
\par

\subsection{Static versus TNE Loops}
By `TNE solutions', we specifically refer to nonsteady solutions in which the plasma repeatedly undergoes cycles of 
(i) a heating phase that results in a nearly flat temperature profile in the corona; 
(ii) the temperature slowly decreases with a dip developing at the apex; 
(iii) the dip evolving into a condensation; 
(iv) the condensation falling to one side under gravity, evacuating the entire loop, and disappearing into the chromosphere; 
(v) the loop refilling with evaporated plasma as it reheats, at which point the loop returns to step (i). 
This cyclical behavior was observed in early numerical studies \citep{1982A&A...108L...1K,1983A&A...123..216M, 1991ApJ...378..372A,1999ApJ...512..985A}, and it has since been independently confirmed in numerous investigations \citep[e.g.][]{2013ApJ...773...94M, 2015ApJ...806..115K, 2016ApJ...827..152A, 2018ApJ...853..176A,2020A&A...633A..11F,2024ApJ...973L...1L}.
\par

TNE solutions spend most of their time in step (ii) of the cycle, and the solution profiles closely resemble static solutions during this phase. 
Note that in the simulations with perfect symmetry, the condensation sits at the apex for a long time. 
This is not the case with any realistic asymmetries.
We first carry out a phase diagram analysis of the static solution (Loop 1).  
Temperature and density `tracks' of the solution are plotted as colored points in the $T$-$n$ phase plane in the top panel of Fig.~\ref{fig:fp_TC}.
The tracks for the footpoints of the loop (gray points, at chromospheric temperatures) trace out what is obviously, upon referring back to Fig.~\ref{fig:spex_Lc}, the lower branch of an equilibrium curve, while the transition region resembles the isobaric path plotted in Fig.~\ref{fig:spex_Lc} that occupies nonequilibrium parameter space.   
This is as we expect: the chromospheric plasma is dense and nearly isothermal, while thermal conduction acts as a heating mechanism in the transition region, serving to raise the plasma to high temperatures.  
\par

The apex point of the loop and a position approximately $4 \times 10^9$ cm lower, referred to as the coronal edge, are marked by the red and blue points (and correspond to the red and blue points in the lower panels of Fig.~\ref{fig:fp_TC}).  
Overplotted are equilibrium curves calculated at both the apex (red line) and the coronal edge (blue line).
Consistent with above, these curves connect to the tracks traced out by the chromospheric plasma.
The yellow band between these two curves illustrates the range of equilibrium curves at each intermediate location between the red and blue points, while the gray points lie outside of these apex and edge positions.
Notice that both the apex and edge points lie above their respective equilibrium curves, indicating that the entire corona is hotter than the temperature where radiative cooling balances heating. 
\par

Is the parameter space above the equilibrium curves a region of net heating or net cooling?  
This is a question about both energy balance and stability that epitomizes the subtleties involved in thinking about the role of thermal conduction.  
We are accustomed to associating steady state solutions with stable equilibrium curves, in the sense that plasma displaced above (below) the equilibrium curve will cool (heat) back toward equilibrium.  
That intuition would lead us to believe the corona occupies a region of net cooling.
However, the cooling function adopted by \citetalias{2019ApJ...884...68K}—like the SPEX function—produces a phase diagram in which nearly all regions at coronal temperatures (above a few $10^5~\rm{K}$) and lying above the equilibrium curve correspond to net heating.
In both the isobaric and isochoric sense, these regions would be thermally unstable—were it not for the strong stabilizing influence of thermal conduction.
The energy balance actually requires that the coronal plasma settle into a heating region because thermal conduction is acting to cool the corona: by \eqref{eq:energy}, steady state solutions satisfy
\begin{equation}
    \nabla \cdot (T^{5/2} \mathbf{\nabla} T) = \frac{\rho}{\chi} \mathcal{L}.
    \label{eq:TCbalance}
\end{equation}
At the apex where $\nabla T = 0$ and the temperature is a maximum, it must be that $\nabla^2 T < 0$, in turn requiring $\mathcal{L} < 0$ (i.e. net heating).  
\par

It becomes inconvenient to assess linear stability using phase diagrams for two reasons.
First, similarly to the equilibrium curve ``bands'' in Fig.~\ref{fig:fp_TC}, TI zone boundaries also become bands unless analyzed point-by-point. 
Second, the thermal conduction term depends on the perturbation wavelength, and as Fig.~\ref{fig:TC} shows, this creates a range of TI zone boundaries even for a point-by-point analysis.
However, it is quite straight forward to perform a linear stability analysis on $\Lambda(T)$ vs. $s$ plots, as shown below.
\par

\subsection{TI mode stability analysis}
In studies adopting the TNE heating prescription to model coronal loops, TI has only been assessed by examining Field's instability criteria \citep[i.e.][]{2015AdSpR..56.2738M, 2020A&A...636A.112C, 2023SoPh..298..102M, 2024ApJ...976..226S, 2024ApJ...973L...1L}. 
As already emphasized in \S{3}, Field's criteria do not apply off the equilibrium curve where the dynamics of TNE takes place.
Indeed, as we just showed, even the static loop has $\mathcal{L} \neq 0$ according to \eqref{eq:TCbalance}. 
\par

\begin{figure}[ht!]
\centering
\plotone{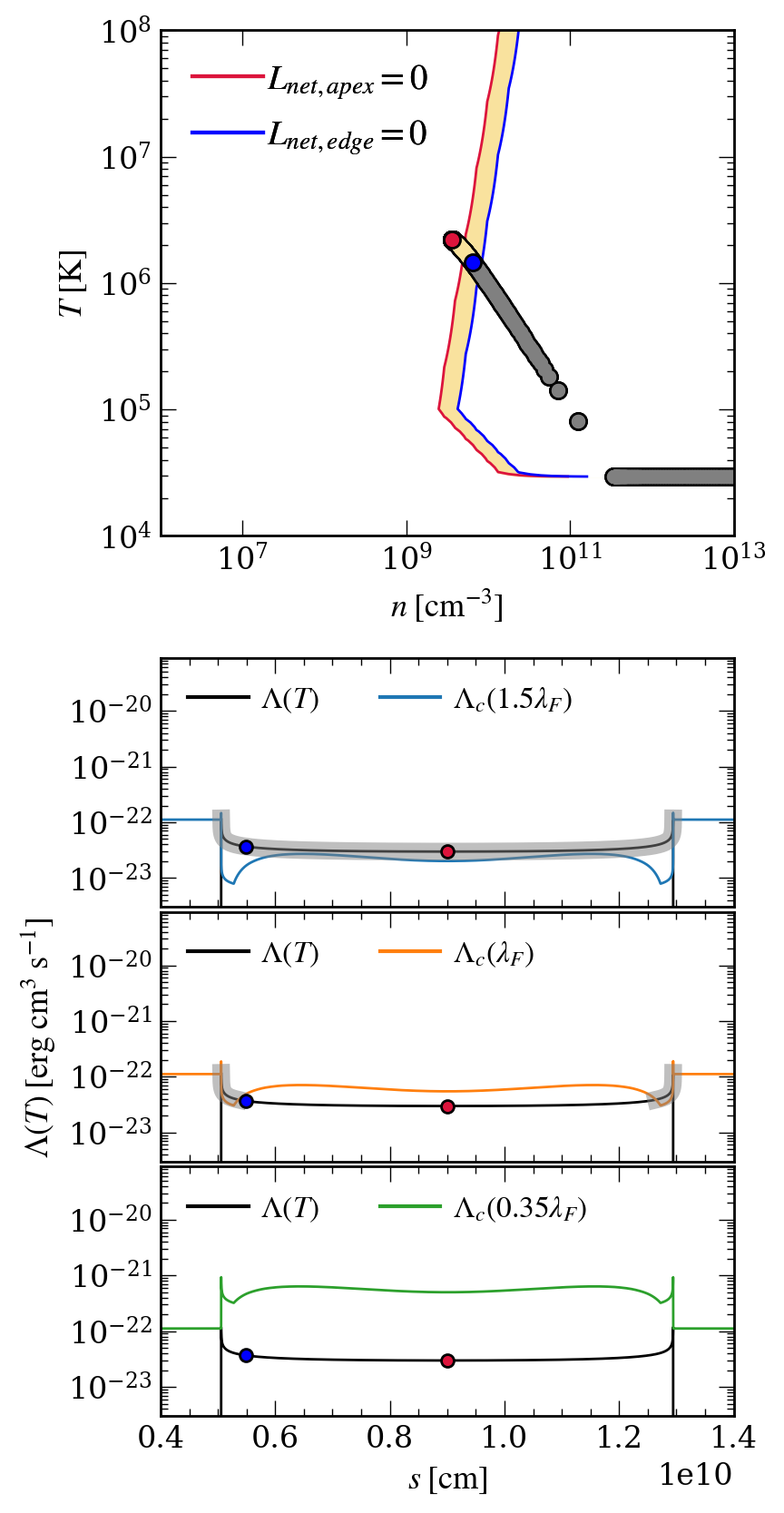}
\caption{Stability analysis of a stable coronal loop simulation from \citetalias{2019ApJ...884...68K}.
\textit{Top panel:} `Tracks' of the steady state solution (Loop 1) are plotted as gray points on a $T-n$ phase diagram; the apex and the coronal edge position are marked with blue and red dots.
The blue and red lines correspond to equilibrium curves computed at the blue and red dots (which are also marked in the lower panels). 
The grey band illustrates the range of equilibrium curves at each intermediate location between these points.   
\textit{Lower panels:} The cooling rate, $\Lambda(T)$, and the critical isobaric cooling rate as functions of the loop position for $\lambda = 1.5\lambda_F$, $\lambda_F$, and $0.35\lambda_F$.}
\label{fig:fp_TC}
\end{figure}

Before we can apply a linear stability analysis based on Balbus' criteria, we need to establish the validity of the local approximation, which requires the gradient scale height in a region to greatly exceed the unstable perturbation wavelengths.
The longest relevant wavelength is the length of coronal portion of the simulation data, $L_{\rm{corona}} \approx 8 \times 10^{9}$ cm.
We take as the characteristic gradient scale height
\begin{equation}
    L_T = \frac{T}{|d T/d s|},
\end{equation}
where $s$ represents the spatial coordinate along the loop.
Computing numerical derivatives at the apex of the loop\textemdash where gradients in temperature and density approach zero\textemdash can lead to artificially inflated scale heights. 
To mitigate this, we evaluate this derivatives at a lower position along the loop that remains well within coronal temperatures.
Specifically, evaluating the numerical derivative at a point a quarter the way down from the blue to the red dots in Fig.~\ref{fig:fp_TC} gives $L_T = 6.8 \times 10^{10}$ cm, which is roughly 8.6 times greater than $L_{\rm{corona}}$. Thus, we are close enough to having $L_T \gg L_{\rm{corona}}$ that linear theory should be applicable.
\par

Only perturbations with wavelengths satisfying $\lambda < L_{\rm{corona}}$ have any meaning. 
By \eqref{eq:field_l2}, the Field length for Loop~2 is $\lambda_F = 3 \times 10^{10}$ cm.  
Because $\lambda_F > L_{\rm{corona}}$, we expect all TI modes excited in the corona to be stabilized by thermal conduction.  
Since we have confirmed this using data for Loop~2, it suffices to use Loop~1 for our analysis.
\par

In the bottom panels of Fig.~\ref{fig:fp_TC}, rather than comparing the loss function with $\Lambda_c$ using cooling curve plots as in Fig.~\ref{fig:TC}, we plot these quantities as a function of distance along Loop~1.
The thermal conduction term is no longer a free parameter as it was in \S{3.3}, so by reducing $\lambda_F$, we are mimicking what would occur in larger or cooler loops. 
In the bottom panel of Fig.~\ref{fig:fp_TC}, we evaluate $\Lambda_c$ at $\lambda = L_{\rm{corona}}$, which corresponds to $\lambda = 0.35\lambda_F$.  
Clearly, the loop is stable by a wide margin.
In the next panel up, we decrease the thermal conduction term by a factor of a few by setting $\lambda = \lambda_F$.  The footpoints of the loop now satisfy $\Lambda(T) > \Lambda_c$.
Only by increasing the wavelength to $\lambda = 1.5\lambda_F$ do we see the entire coronal region of the     loop enter the isobaric TI zone (as shown by the next panel up).
\par 

\subsection{CC mode stability analysis}
\begin{figure*}[ht!]
\centering
\begin{interactive}{animation}{figure8animation.mp4}
\plotone{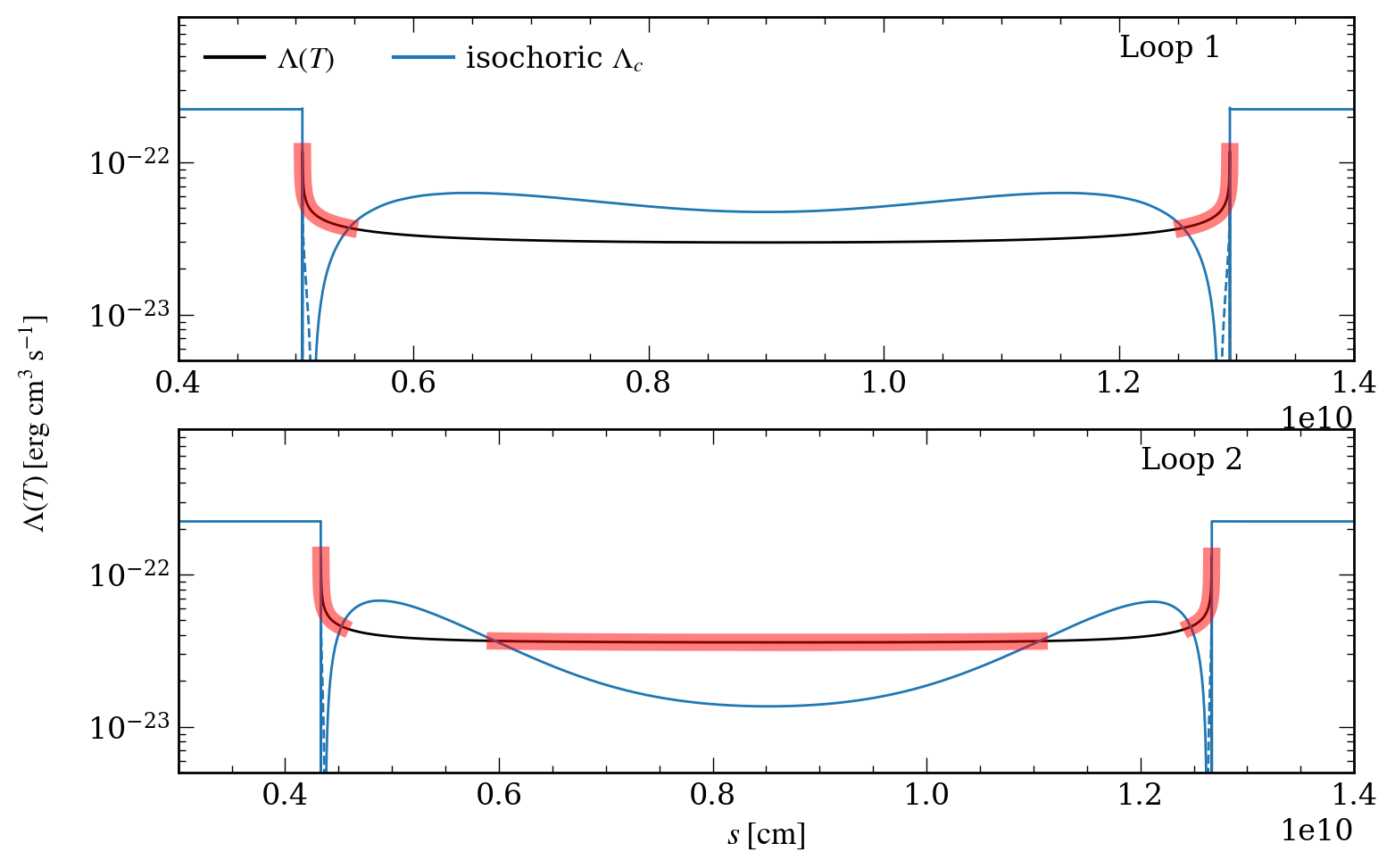}
\end{interactive}
\caption{Cooling curve based stability analysis of a stable solution (top panel) and a TNE solution (bottom panel) from \citetalias{2019ApJ...884...68K}.
Black lines are $\Lambda(T)$, while blue lines are the isochoric $\Lambda_c$ values from \eqref{eq:cc_mode_Lambdac}; red highlighted portions of the black lines indicate where $\Lambda(T) > \Lambda_c$.
Loop 2 is shown just after the temperature reaches its maximum during the TNE cycle and is unstable to CC modes, yet stable to TI modes.
An animation of the bottom panel can be found in the online journal. The animation shows the evolution of $\Lambda$ and isochoric $\Lambda_c$ for Loop 2. The coronal portion loop begins in the stable regime ($\Lambda(T) < \Lambda_c$), with the red highlighted portions of the black lines indicate where $\Lambda(T) > \Lambda_c$. The bottom panel of the figure shows the state of the loop 1 hour later once $\Lambda(T)$ (black line) has begun to exceed the critical cooling rate, isochoric $\Lambda_c$ (blue line). A dense, cool condensation forms within the loop around 3.6 hours later, which is comparable to the isochoric growth time.}
\label{fig:loop_comp}
\end{figure*}

Despite our derivation in \citetalias{2025SoPh..300....5W} closely following that of \citet{1953ApJ...117..431P}, whose paper introduced the very concept of TI, we showed that Field's isochoric instability criterion governs, in addition to isochoric TI modes, the mode of what should be considered a separate instability.\footnote{Immediately after the publication of \citetalias{2025SoPh..300....5W}, Ramon Oliver, Jaume Terradas, and Varsha Felsyand pointed out to us a flaw in the logic of that papers' Appendix~B where we claimed the CC mode is not the $k=0$ limit of an isochoric TI mode.} 
That is, if one takes the operational definition of an unstable TI mode to be a sinusoidal perturbation undergoing exponential amplification of both the underdense and overdense portions, then this definition excludes a `pure' isochoric mode, i.e. a temperature fluctuation having $\delta \rho = 0$.  
We named this the catastrophic cooling (CC) mode because it acts to exponentially amplify the quantity
\begin{equation}
    \delta T = \avg{T(x)} - T_{\rm{eq}},
    \label{eq:cc_mode}
\end{equation}
which is the difference between the mean gas temperature $\avg{T(x)}$ and $T_{eq}$, the equilibrium temperature satisfying $\mathcal{L}(T_{eq}) = 0$.
Conversely, an isochorically stable plasma is one where this mode is being damped, i.e. the CC mode then serves as the agent driving the system toward thermal equilibrium.  
\par

Another way to reach the conclusion that catastrophic cooling instability is in a category of its own is to recognize that because TI recovers this instability in the $k=0$ limit, it will occur in ``0D'' as well as in 1D, 2D, and 3D simulations.  Stated differently, our derivation in \citetalias{2025SoPh..300....5W} is based on solving a time-dependent ordinary differential equation rather than partial differential equations.
Not only can the CC mode compete with and even dominate all TI modes, it is the fastest growing mode in the system whenever $0<R<1$, where $R = N_p^\prime/\gamma N_\rho^\prime$ is the basic parameter characterizing different regimes of TI (see \citetalias{2025SoPh..300....5W}).

Crucial to explaining the onset of catastrophic cooling in TNE solutions is the fact that thermal conduction cannot stabilize CC modes (provided the local approximation holds).
This is because $\delta T$ is spatially constant by \eqref{eq:cc_mode} and so $\nabla^2 \delta T = 0$.  Therefore, upon setting $\lambda_F = 0$ in \eqref{eq:isochor_L}, we are left with
\begin{equation}
    \label{eq:cc_mode_Lambdac}
    \textnormal{(isochoric)} \; \Lambda_c = \Gamma(n,s) + T\frac{d\Lambda(T)}{dT}.
\end{equation}
We should note that if the local approximation does break down, the neglected background temperature gradients contribute both stabilizing and destabilizing effects to CC modes.\footnote{By applying the  Eulerian perturbation operator $\delta$ to the thermal conduction term in \eqref{eq:energy}, we have $\nabla \cdot [\delta\left(T^{5/2} \nabla T \right)] = \frac{15}{4} T^{1/2} \delta T (\nabla T)^2 + \frac{5}{2} T^{3/2} \delta T \nabla^2 T + 5 T^{3/2} \nabla T \cdot \nabla (\delta T) + T^{5/2} \nabla^2 (\delta T)$. 
Since $\delta T$ is a constant for CC modes, the last two terms are zero and the expression becomes $\nabla \cdot [\delta\left(T^{5/2} \nabla T \right)] = \frac{5}{2} T^{3/2} \delta T [\frac{3}{2} (\nabla T)^2/T + \nabla^2 T$].
Because the sign of the first bracketed term is positive, this term is always destabilizing for CC modes (which by definition have $\delta T < 0$), i.e. it leads to further cooling.  The sign of the second term varies with time and will be stabilizing for CC modes when $\nabla^2 T < 0$, i.e. before a dip in temperature develops.}
\par 

In Fig.~\ref{fig:loop_comp}, we plot both $\Lambda(T)$ and isochoric $\Lambda_c$ along Loop~1 (top panel) and Loop~2 (bottom panel) to assess the stability of these loops to CC modes.  
The regions of each loop residing in the isochoric TI zone are highlighted in red. 
For Loop~1, the entire corona is stable, and we see that only the transition region enters the isochoric TI zone. 
(Because the transition region is host to strong temperature gradients, the local approximation is not valid, making linear theory results ambiguous.)
In contrast, the bottom panel shows that for Loop~2, a majority of the coronal region of the loop occupies the isochoric TI zone.  
This occurs at the time of peak temperature during one of several TNE cycles, approximately 3.6 hours before the cold condensation forms. 
We note that the lower $Q_f$ in the footpoints of Loop~2, along with a smaller heating scale height, results in less heating in the corona where the condensation forms.  
This reduced heating lowers the threshold to instability (i.e. lowers $\Gamma(n)$ in \eqref{eq:cc_mode_Lambdac}), and it should also be noted that any change in heating parameters will introduce CC modes into the system.
\par

An animation accompanying the bottom panel of Fig.~\ref{fig:loop_comp} can be found in the online journal.
It illustrates the evolution of isochoric $\Lambda_c$ starting from the peak temperature point in a TNE cycle, approximately 13 thermal times (11.8 hours) after exponential heating began and the loop had stabilized into recurring TNE cycles.
Loop~2 starts off stable, and as time evolves, $\Lambda(T)$ can be seen rising above the critical cooling rate. 
A condensation forms about 3.6 hours later, and this can be compared with the timescale associated with the isochoric growth rate, $t_{th}/N_\rho^\prime = 6.6$ hours.  
Thus, our analysis points to the cause of condensations in TNE loops being the exponential growth of CC modes.
\par

\section{Summary \& Conclusions}
We have presented the following results:
\begin{itemize}
    \item Balbus' instability criteria for TI are equivalent to a lower bound on the cooling rate, $\Lambda(T) > \Lambda_c$, where there are separate values of $\Lambda_c$ for isochoric and isobaric TI modes.  CC modes are unstable if $\Lambda(T)$ exceeds the $\Lambda_c$ value given by \eqref{eq:cc_mode_Lambdac}.
    \item The critical value $\Lambda(T) = \Lambda_c$ defines the boundary of TI zones, revealing the full unstable parameter space accessible to gas in nonequilibrium states.
    \item Stabilization of TI modes by thermal conduction can be quantified using a $\Lambda_c$-based analysis and visualized using either phase diagrams or $\Lambda(T)$-vs-$T$ plots. 
    \item A $\Lambda_c$-based analysis of the global 1D coronal loop simulations performed by \citetalias{2019ApJ...884...68K} (using a standard TNE prescription) reveals that the non-TNE loop is linearly stable to both TI modes and CC modes, while the TNE loop is unstable only to CC modes.
\end{itemize}
\par

The last result is also noted in the final columns of Table~\ref{table1}. 
We emphasize this result as we think it resolves a longstanding debate in the solar physics community, namely the origin of the catastrophic cooling process taking place in TNE solutions.
One side argues that TNE itself is a nonlinear condensation formation process \citep[e.g.][]{2004A&A...424..289M, 2010ApJ...716..154A, 2010ApJ...717..163B, 2015ApJ...807..158F, 2017ApJ...835..272F, 2018ApJ...855...52F, 2019SoPh..294..173K}, while another ultimately attributes condensation formation to TI \citep[e.g.][]{2011ApJ...737...27X,2013ApJ...771L..29F,2015ApJ...807..142F, 2015AdSpR..56.2738M, 2019A&A...625A.149J,2022ApJ...926L..29A,2023MNRAS.526.1646D,2024ApJ...971...90D}.
\par 

In \S{4} we showed that a careful stability analysis supports an alternative possibility: catastrophic cooling instability, the simple yet subtle linear instability that TI reduces to in the infinite wavelength limit, is being triggered locally (potentially zone-by-zone) prior to the onset of condensation formation in TNE simulations.  
\par

If our conclusions are correct and the debate over the cause of condensations in TNE loops can be settled, the resolution to the debate and the revised explanation for the origin of catastrophic cooling in TNE solutions is as follows.
Before this work, the application of linear theory to coronal loop simulations was limited to examining Field's criteria, and when cast in terms of $L_{net}$ (see \eqref{eq:lnet}), his isochoric criterion reads
\begin{equation}
    \frac{d \Lambda(T)}{d T} < 0.
    \label{eq:field_isochoric}
\end{equation}
Here we have dropped the stabilizing contribution of a thermal conduction term to this criterion because it is zero for the CC mode (although see Footnote~6) and hence not relevant to our explanation of the behavior of TNE solutions. 
Nevertheless, in \S{4.3} we emphasized the importance of this term for ruling out TI modes as a possible cause of catastrophic cooling; again, these modes are stabilized anytime the Field length exceeds the loop length (the typical situation).  
Balbus' isochoric instability criterion, or equivalently our criterion $\Lambda(T) > \Lambda_c$ with $\Lambda_c$ given by \eqref{eq:cc_mode_Lambdac}, is
\begin{equation}
     \Lambda(T) > \Gamma(n) + T \frac{d\Lambda(T)}{dT}. 
     \label{eq:balbus_isochoric}
\end{equation}
We see that the inequalities in \eqref{eq:field_isochoric} and \eqref{eq:balbus_isochoric} are equivalent for equilibrium states with $\Lambda(T) = \Gamma(n)$.  
However, for nonequilibrium states, the heating term in \eqref{eq:balbus_isochoric} is what sets the threshold for catastrophic cooling instability. 
In stable loops with more uniformly distributed heating, $\Gamma(n)$, which is proportional to $n^{-2}$ under the most commonly adopted TNE heating prescription, is large in the tenuous corona, making the threshold to instability large.  In TNE loops, the exponential decline in footpoint heating is steeper because the heating scale height is smaller (see Table~\ref{table1}).  
Thus, $\Gamma(n)$ becomes smaller within the corona and the instability threshold is easily overcome.
Despite the simplicity of this explanation, it appears to successfully summarize the necessary condition for obtaining TNE solutions versus static ones. 
\par

A caveat to this conclusion is that while our analysis shows that Loop~2 is unstable to a local linear instability, it has not ruled out a possible role for a global linear instability. 
A recent study by \citet{2025arXiv250623591K} finds no global eigenmodes in the isochoric limit for semi-circular coronal loop configurations, suggesting that the CC mode may not have a valid global counterpart. 
However, their result stands in direct conflict with the earlier analysis by \citet{1985ApJ...291..798B}, who identified an unstable $k=0$ global mode, also for a semi-circular loop.
\citet{1985ApJ...291..798B}'s $k=0$ unstable eigenmode is free of `nodes', while all eigenmodes with one or more nodes are stable, a situation exactly analogous to all TI modes being stable and the CC mode unstable.
Until it is shown that this earlier work is flawed, the absence of an unstable $k=0$ global mode according to \citet{2025arXiv250623591K}'s analysis should be regarded with skepticism.
\par

Furthermore, our analysis has not ruled out a nonlinear process for condensation formation that will occur in the absence of any instability, and TNE has been interpreted as such a process \citep{2019SoPh..294..173K}.  
In general, such a condensation process will involve velocity gradients, as evaporative upflows carrying substantial enthalpy are present during both the heating and cooling phases in TNE loops \citepalias{2019ApJ...884...68K}. 
A stability analysis must then account for the heating contribution from the divergence of the enthalpy flux \citep{2010ApJ...717..163B}.
\par

More work is therefore needed to definitively establish catastrophic cooling instability as the underlying condensation mechanism.  
Aside from the above caveats, this entails closing a loophole that we have so far neglected to mention: condensation growth occurs nearly isobarically in Loop~2, while the CC mode can only grow isochorically in local simulations.
This is clearly visible in the pressure profile animation\footnote{The animation is provided in the online journal and at \url{https://astricklan.github.io/SWK2025/.}} accompanying Figure~\ref{fig:profile}.
This discrepancy is likely due to how a local instability saturates in a global simulation.
That is, upon evolving at constant density within each zone throughout the linear regime, the amount of pressure support being lost would likely be minimal enough to not show up in the pressure profile on the scale of the full loop.  
Once the amplitude grows to be nonlinear, however, the background plasma would quickly remedy any pressure imbalance, making it appear as though an isobaric condensation process is taking place.
\par

\begin{acknowledgments}
AS thanks Munan Gong for enlightening discussions on computing heating and cooling rates from chemistry networks.
TW acknowledges the DOE Office of Science's \href{https://science.osti.gov/wdts/suli}{SULI program}, 
through which Tess Boland was employed at LANL and wrote several analysis routines using SymPy for a related project that we built upon for this study. 
JK thanks M. Luna for providing the equilibrium loop solution.

This work was supported by the U.S. Department of Energy through the Los Alamos National Laboratory. 
Los Alamos National Laboratory is operated by Triad National Security, LLC, for the National Nuclear Security Administration of U.S. Department of Energy (Contract No. 89233218CNA000001). 
JK was supported by the Internal Scientist Funding Model (competitive grant program) at GSFC. 
He benefited from participation in the International Space Science Institute team ``Observe Local Think Global: What Solar Observations Can Teach Us About Multiphase Plasmas Across Astrophysical Scales" led by P. Antolin and C. Froment.
\end{acknowledgments}

\clearpage

\appendix
\section{Radiative Loss Functions}
\label{app:1}
\subsection{Koyama \& Inutsuka (2002) (KI)}
The interstellar medium (ISM) radiative loss function was first presented by \citet{2002ApJ...564L..97K}.
It was noted by \cite{2007ApJ...657..870V} that Eq. (4) in \citet{2002ApJ...564L..97K} contains two typographical errors. We include here the corrections provided by Koyama and Inutsuka to Vaz\'quez-Semadeni.
\begin{equation}
    \label{eq:koyama}
    \Lambda(T) = 2 \times 10^{-26}\left[10^7\exp\left(\frac{-1.184\times 10^5}{T + 1000}\right) + 1.4\times10^{-2}\sqrt{T}\exp\left(\frac{-92}{T}\right) \right] \;\;\; \textnormal{ergs cm$^3$ s$^{-1}$},
\end{equation}

\subsection{Schure et al. (2009) (SPEX)}
The SPEX radiative loss function comes from \citet{Hermans_EPS_2021}\footnote{Obtainable at \url{https://erc-prominent.github.io/team/jorishermans/}}, who fit the tabular form of the SPEX\_DM table into the following piecewise function:
\begin{equation}
    \label{eq:spex}
    \Lambda(T) = 
    \begin{cases}
        10^{-35.314} \ T^{5.452} & \ 10 < T \leq 10^{1.422} \text{K} \\
        10^{-29.195} \ T^{1.150} & \ 10^{1.422} < T \leq 10^{2.806} \text{K} \\
        10^{-26.912} \ T^{0.337} & \ 10^{2.806} < T \leq 10^{3.980} \text{K} \\
        10^{-108.273} \ T^{20.777} & \ 10^{3.980} < T \leq 10^{4.177} \text{K} \\
        10^{-18.971} \ T^{-0.602} & \ 10^{4.177} < T \leq 10^{4.443} \text{K} \\
        10^{-32.195} \ T^{2.374} & \ 10^{4.443} < T \leq 10^{4.832} \text{K} \\
        10^{-21.217} \ T^{0.102} & \ 10^{4.832} < T \leq 10^{5.397} \text{K} \\
        10^{-0.247} \ T^{-3.784} & \ 10^{5.397} < T \leq 10^{5.570} \text{K} \\
        10^{-15.415} \ T^{-1.061} & \ 10^{5.570} < T \leq 10^{5.890} \text{K} \\
        10^{-19.275} \ T^{-0.406} & \ 10^{5.890} < T \leq 10^{6.232} \text{K} \\
        10^{-9.387} \ T^{-1.992} & \ 10^{6.232} < T \leq 10^{6.505} \text{K} \\
        10^{-22.476} \ T^{0.020} & \ 10^{6.505} < T \leq 10^{6.941} \text{K} \\
        10^{-17.437} \ T^{-0.706} & \ 10^{6.941} < T \leq 10^{7.385} \text{K} \\
        10^{-25.026} \ T^{0.321} & \ 10^{7.385} < T \leq 10^{8.160} \text{K} \\
    \end{cases}
\end{equation}
The SPEX table was calculated with the SPEX package \citep{1996uxsa.conf..411K} using solar abundances from \citet{1989GeCoA..53..197A} and implemented in the AMRVAC MHD code \citep{2003CoPhC.153..317K}.
This function is representative of many other coronal cooling functions used in the last couple decades \citep[e.g.][]{2008ApJ...689..585C, 2011piim.book.....D, 2016A&A...587A.151K, 2021A&A...655A..36H}.
\par

\subsection{Klimchuk \& Luna (2019)}
The \citetalias{2019ApJ...884...68K} radiative loss function is a simplified piecewise function used in the 1D coronal loop simulations analyzed in \S{4}.
We note their use of $n_e$ rather than the total number density in \eqref{eq:rhoL}, as \citetalias{2019ApJ...884...68K} assume a fully ionized plasma ($\bar{m} = 0.5 m_p$).  The factor of $1/4$ here accounts for this difference:
\begin{equation}
    \Lambda(T) = \frac{1}{4} 
    \begin{cases}
        10^{-18.75} \ T^{-1/2} & \ \;\;\;\;\;\;\; T \geq 10^{5} \text{K} \\
        10^{-36.25} \ T^{3} & \ 3 \times 10^{4} \leq T < 10^{5} \text{K} \\
        10^{-25.518} \ (T - 2.95 \times 10^{4}) & \ 2.95 \times 10^{4} \leq T < 3 \times 10^{4} \text{K} \\
    \end{cases}
\end{equation}

\section{Critical cooling rates with arbitrary density dependence}
\label{app:2}
Here we repeat the derivation given in \S{3.1} assuming that $\Lambda = \Lambda(T,n)$ and $\Gamma = \Gamma(T,n)$ instead of $\Lambda = \Lambda(T)$ and $\Gamma = \Gamma(n)$.
Again, starting with the instability criteria in \eqref{eq:Np} and \eqref{eq:Nrho}, we substitute in \eqref{eq:Np_dim} and \eqref{eq:Nrho_dim} and then eliminate $L_{net}$ using \eqref{eq:lnet} to arrive at 
\begin{align}
\left( \frac{\partial \Lambda(T,n)}{\partial T} \right)_n - \left( \frac{\partial\Gamma(T,n)}{\partial T} \right)_n  - \frac{n}{T}\left( \frac{\partial\Lambda(T,n)}{\partial n} \right)_T+ \frac{n}{T}\left( \frac{\partial \Gamma(T,n)}{\partial n} \right)_T- \frac{2\Lambda(T,n)}{T} + \frac{2\Gamma(T,n)}{T} &< - \frac{\Em}{T}\left(\frac{\lambda_F}{\lambda}\right)^2 ;\\
 \left( \frac{\partial\Lambda(T,n)}{\partial T} \right)_n - \left( \frac{\partial\Gamma(T,n)}{\partial T} \right)_n - \frac{\Lambda(T,n)}{T} + \frac{\Gamma(T,n)}{T} &< - \frac{\Em}{T}\left(\frac{\lambda_F}{\lambda}\right)^2 .
\end{align}
Solving for $\Lambda(T,n)$ as before, the full $\Lambda_c$ definitions read
\begin{align}
(\textnormal{isobaric}) \; \Lambda_c &= \Gamma(T,n) + \frac{1}{2}\bigg[T\left( \frac{\partial\Lambda(T,n)}{\partial T} \right)_n - T\left( \frac{\partial\Gamma(T,n)}{\partial T} \right)_n - n\left( \frac{\partial\Lambda(T,n)}{\partial n} \right)_T + n\left( \frac{\partial\Gamma(T,n)}{\partial n} \right)_T + \Em\left(\frac{\lambda_F}{\lambda}\right)^2 \bigg] ; \\
(\textnormal{isochoric}) \; \Lambda_c &=\Gamma(T,n) + T\left( \frac{\partial\Lambda(T,n)}{\partial T} \right)_n - T\left( \frac{\partial\Gamma(T,n)}{\partial T} \right)_n + \Em\left(\frac{\lambda_F}{\lambda}\right)^2.
\end{align}
\par

All spatial derivatives appearing in $\Lambda_c$ are evaluated using a custom python code developed for this study using the differentiation capabilities of SymPy \citep{10.7717/peerj-cs.103}. 
The temperature and density profiles are first represented as symbolic variables using \texttt{sympy.symbols} and passed into the specified radiative loss function. 
SymPy then computes the necessary derivatives symbolically, with \texttt{sympy.diff}, as one would by hand. 
To evaluate the radiative loss functions given by \eqref{eq:koyama} and \eqref{eq:spex}, SymPy automatically selects and differentiates the appropriate segment corresponding to the specified temperature range.
\par

\clearpage

\section{Pressure profile animation}
\label{app:3}

\begin{figure*}[h!]
\centering
\begin{interactive}{animation}{figure9animation.mp4}
\includegraphics[width=\textwidth]{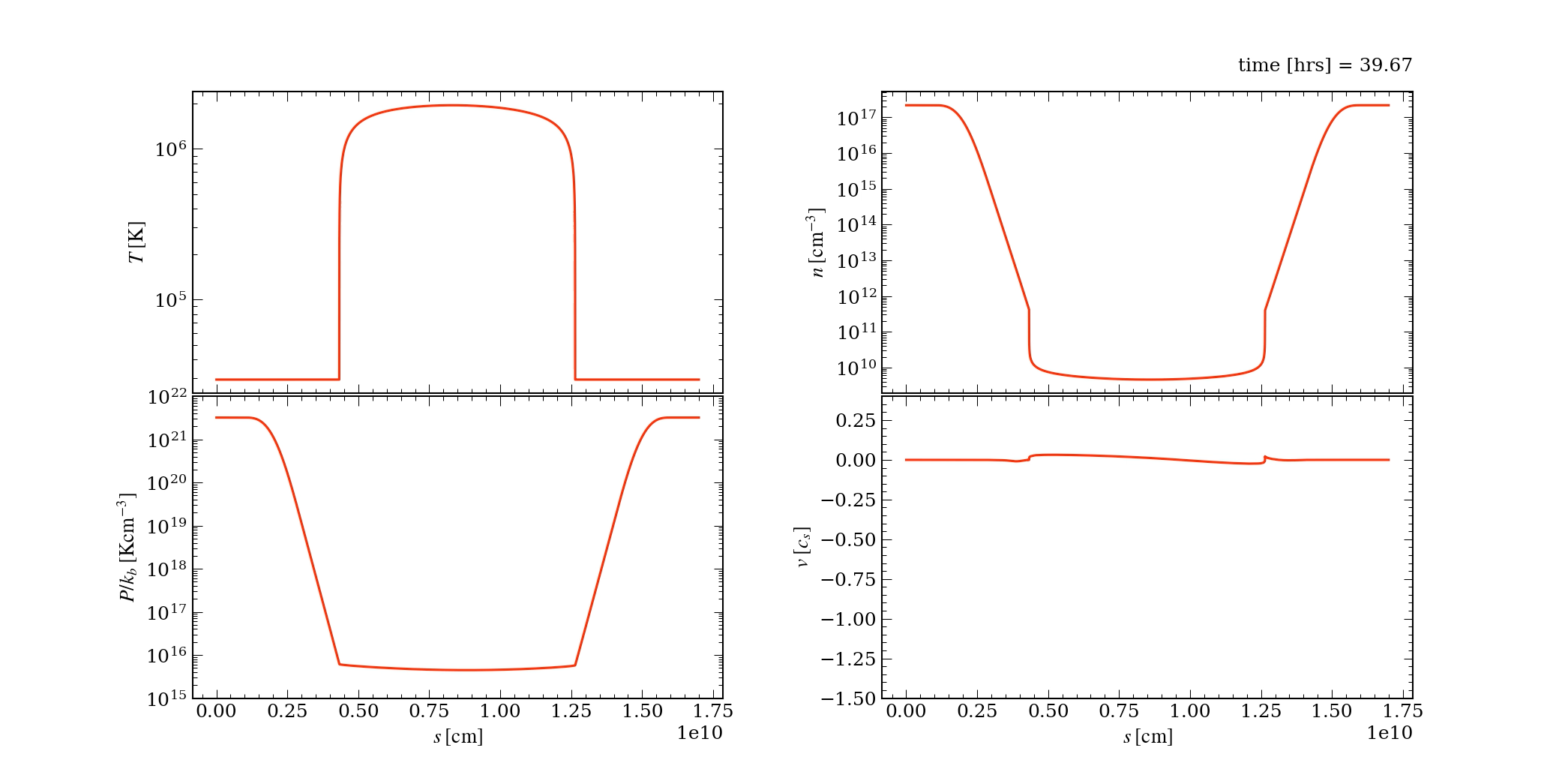}
\end{interactive}
\caption{An animation of Loop~2 temperature, density, pressure, and velocity profiles. The first frame begins from a state where the loop has stabilized into a repeating thermal nonequilibrium (TNE) cycle. The first time step is at peak temperature during one TNE cycle, about 11.8 hours (13 thermal times) after exponential heating was fully implemented in the simulation. As the condensation forms, note the apparent isobaric evolution. This discrepancy likely reflects how a local instability saturates within the context of a global simulation. Each zone appears to evolve at nearly constant density during the linear phase, potentially resulting in only minor local pressure loss--perhaps too subtle to register in the full-loop pressure profile. As the perturbation amplitude grows and nonlinear effects set in, we hypothesize that the surrounding plasma dynamically compensates for emerging pressure imbalances. This adjustment could give the impression of an isobaric condensation process, even though the underlying instability originated under isochoric conditions.}
\label{fig:profile}
\end{figure*}

\clearpage

\bibliographystyle{aasjournal}
\bibliography{references.bib}

\end{document}